\documentclass[journal]{IEEEtran}
\usepackage[utf8]{inputenc}
\usepackage[T1]{fontenc}
\usepackage[strings]{underscore}
\usepackage{graphicx}
\usepackage{booktabs}
\usepackage{amsmath,amssymb}
\usepackage{array}
\usepackage{tabularx}
\usepackage{multirow}
\usepackage{xcolor}
\usepackage{flushend}
\usepackage[font=footnotesize,labelfont=bf]{caption}
\usepackage{enumitem}
\setlist{nosep,leftmargin=1.3em}
\usepackage{listings}
\usepackage{textcomp}

\definecolor{codebg}{HTML}{F6F8FA}
\definecolor{navy}{HTML}{1E3A5F}
\definecolor{kw}{HTML}{1E40AF}
\definecolor{cm}{HTML}{6B7280}
\definecolor{st}{HTML}{B45309}

\lstdefinelanguage{SV}{
  morekeywords={module,endmodule,input,output,logic,wire,reg,always_ff,always_comb,
    posedge,negedge,if,else,begin,end,case,endcase,unique,typedef,enum,struct,packed,
    assign,localparam,parameter,for,int,bit,byte,signed,unsigned,function,endfunction,
    return,default,genvar,generate,endgenerate,always,initial,longint,shortint,break,
    automatic,import,package,endpackage,localparam},
  sensitive=true, morecomment=[l]{//}, morecomment=[s]{/*}{*/}, morestring=[b]"}
\usepackage{hyperref}
\hypersetup{colorlinks=true,linkcolor=navy,citecolor=navy,urlcolor=navy}
\graphicspath{{../figs/}}

\newcommand{\code}[1]{\texttt{\small #1}}

\begin{document}

\title{DTX: A Throughput-First Training Accelerator for Diffusion and
Transformer Models}

\author{Shashank, Independent Researcher, SF, CA%
\thanks{ Corresponding contact:
\texttt{sshashan@alumni.usc.edu}.}}

\markboth{Shashank: Throughput-First Training Accelerator (DTX) Technical Report, v1.0}%
{Shashank: A Throughput-First Training Accelerator for Diffusion and Transformer Models}

\maketitle

\begin{abstract}
The throughput of a floating-point training accelerator is set by how its
reductions are ordered: any summation serialized through a single FP32
adder is a loop-carried dependence that pins a machine near 2~FLOP per
cycle regardless of physical design. DTX is a training accelerator built
so that no such chain exists anywhere. Every reduction is a
pipelined binary tree, every floating-point operator a two-stage pipeline
with initiation interval~1, and \emph{loop-carried anything is a design bug}.
An $8{\times}8$ weight-stationary systolic array serves all three training
GEMM dataflows with a fused bias/activation/cast drain epilogue; an 8-lane
vector unit covers the twelve non-GEMM training ops; an 8-lane fused AdamW
pipeline retires one element per lane per cycle; and a pipelined
Philox-4$\times$32-10 source streams Gaussians. A 256-bit, 4-slot VLIW word
with compiler-scheduled \code{wait\_mask} dependence bits and zero-overhead
loops co-issues the engines over a unified 64\,KB ping-pong tile space and a
256-bit AXI4 DMA, 216~FLOP per cycle, roughly $108\times$ that loop-carried
floor per clock. With no canonical sum order, the verification contract is
tolerance-based: DTX is checked against an FP64 golden model under a per-element
tolerance derived from tree-reduction error analysis (scaled
$1{+}\log_2(K)/8$ in reduction depth $K$), with exact-equality carve-outs,
and the bound is demonstrably tight, a premise-violating program measured
$5{,}340\times$ over budget, while all 17~tests pass, the training
acceptance test alone sweeping $107{,}108$ elements with zero failures. Semantic gates seal the methodology: an
on-device diffusion-MLP training run must reduce its loss (56.4 to 26.0 over
20 steps) and an attention block verifies forward and backward. Counter-level
overlap proof ($9{,}880$ GEMM $+$ $2{,}114$ DMA busy $>$ $10{,}543$ wall
cycles) shows the peak is sustained, and an explicitly analytical iso-node
decomposition, utilization $2.0$--$2.6\times$ times energy per op
$3$--$4\times$, bounds the GPU comparison at $6$--$10\times$
throughput per watt. A sky130 campaign grounds the physical side: the
physical-design configuration passes the same regression (17/17, catching
one hardcoded-parameter stimulus bug), the full chip synthesizes to 2.58M
cells (77\% of cell area in FF-mapped memory on the macro-less open flow),
and the systolic array hardens to router-DRC-clean GDS at a post-route
83.3\,MHz typical, $1.9\times$ the post-route ceiling of an optimized
loop-carried multiply--accumulate baseline hardened on the same node and
flow.
\end{abstract}

\begin{IEEEkeywords}
Training accelerator, systolic array, VLIW, EPIC, diffusion models, DiT,
transformers, backpropagation, AdamW, bfloat16, mixed precision,
floating-point summation, tolerance-based verification, golden model, Philox,
Gaussian random number generation, AXI4, Universal Verification Methodology,
register-transfer-level design.
\end{IEEEkeywords}

\section{Introduction}\label{sec:intro}

The throughput of a floating-point training accelerator is set less by how
many multipliers it instantiates than by how its reductions are
\emph{ordered}. A summation evaluated as a sequential chain through a
single FP32 adder is a loop-carried dependence: element $k{+}1$ cannot
enter the adder until element $k$ has left it, so either the adder is
combinational, and its carry chain caps $f_{\max}$, or the accumulation
runs at an initiation interval equal to the adder depth. Any datapath
built around such a chain sits at a floor of about
2~FLOP per cycle (one multiply feeding one dependent add), and no
physical-design effort moves it, because the ceiling is the \emph{ordering},
not the technology.

This paper presents DTX, a training accelerator
designed so that no such chain exists anywhere: it prioritizes throughput
over reproduction of any canonical sum order. Every reduction is a pipelined binary tree
(\code{dtx\_tree\_add}/\code{dtx\_tree\_max}, one two-stage FP32 operator per
node, initiation interval~1); the GEMM engine is an $8{\times}8$
weight-stationary systolic array whose per-column accumulation is a fixed
pipeline order with no sequential-sum contract; and a hard design rule, %
\emph{loop-carried anything is a design bug}, is enforced by construction:
every FP operator is a two-stage pipeline, no combinational chain is deeper
than one FP operation, and the Newton reciprocal/rsqrt loops are unrolled one
iteration per pipeline segment. Around the array sit a fused drain epilogue
(bias, ReLU/GELU, optional BF16 cast, no extra memory pass), an 8-lane FP32
SIMD vector unit covering the twelve non-GEMM training ops
(LayerNorm and softmax forward/backward, activation backward, loss gradients,
noising, casts, bias reduction), an 8-lane fully fused AdamW pipeline
(one element per lane per cycle), and a pipelined Philox-4$\times$32-10
Gaussian source. Control is a
256-bit, 4-slot VLIW word, one slot each for GEMM, vector, optimizer/RNG,
and DMA, with EPIC-style compiler-scheduled dependence bits
(\code{wait\_mask}), tile-descriptor indirection, and a zero-overhead loop;
data lives in a unified 64\,KB tile space of three ping-pong regions behind a
2-read/1-write-per-region crossbar, fed by a 256-bit AXI4 DMA with 2-D
tile gather/scatter. In the simulated configuration the machine peaks at
216~FLOP per cycle (128 GEMM $+$ 8 vector $+$ 80 optimizer under the plan's
conservative ten-FLOP-per-element AdamW bookkeeping), roughly $108\times$
the loop-carried floor per clock, before any $f_{\max}$ advantage from
the absence of loop-carried arithmetic.

Choosing parallel reduction orders dictates the verification contract, and
we treat that contract as a first-class design artifact rather
than a concession. A tree reduction still has a \emph{fixed} pairing order, %
DTX remains run-to-run deterministic, and its counter-based RNG is checked
bit-exactly at the integer level, but its result provably differs from any
sequential reference by data-dependent rounding, so byte comparison against
an independent model is no longer meaningful. DTX is therefore verified
against an FP64 golden model under a \emph{derived} per-element tolerance,
$|d-r| \le \mathrm{atol} + \mathrm{rtol}\cdot|r|$ with
$\mathrm{rtol}$ scaled as $(1+\log_2(K)/8)$ in the reduction depth $K$,
exact ($===$) comparison retained for the classes where the math is exact
(copies, casts of exact values, integer Philox output, CSR/IRAM readback),
and a statistical moment check on the Gaussian source. The bound is tight
rather than vacuous: when a test program violated its premise by chaining a
computed BF16 tile into a GEMM, the scoreboard measured a $5{,}340\times$
tolerance violation, while all $78{,}292$ tolerance-checked elements of the
compliant training suite pass with zero failures
(Sec.~\ref{sec:numerics}, \ref{sec:verif}).

Finally, we frame the headline comparison honestly. The oft-quoted
``$10\times$ over a GPU'' for fixed-function training silicon is, in this
paper, an \emph{explicitly analytical} decomposition at iso-node:
a utilization ratio of $2.0$--$2.6\times$ (a systolic array sustaining
${\sim}0.9$ on dense training GEMMs, our in-simulation proxy is 93.7\%
GEMM-slot occupancy with overlapped DMA, against published end-to-end
model-FLOPs utilization of $0.35$--$0.46$ for large-transformer GPU
training), multiplied by an energy-per-op ratio of $3$--$4\times$ (no warp
scheduler, per-op instruction supply, or large register file; fused epilogue
and optimizer eliminating elementwise-kernel memory round-trips), for a
product of $6$--$10\times$ throughput per watt, approaching an order of
magnitude, with $10\times$ the favorable end (Sec.~\ref{sec:results}). The
sky130 prototype exists to measure FLOP/cycle/mm$^2$ and per-op energy;
the scale-out numbers are architecture claims and are labeled as such.

\subsection{Contributions}

\begin{itemize}
\item \textbf{A throughput-first training microarchitecture}
(Sec.~\ref{sec:arch}, \ref{sec:uarch}): an $8{\times}8$ weight-stationary
systolic GEMM array serving all three training dataflows
($Y{=}XW^{\mathsf T}$, $dX{=}dY\,W$, $dW \mathrel{+}= dY^{\mathsf T}X$ with
FP32 accumulation onto the output tile) with a fused
bias/activation/cast drain epilogue; an 8-lane vector unit; an 8-lane fused
AdamW datapath with PyTorch-compatible semantics; and a pipelined,
epoch-tagged Philox Gaussian source, 216~FLOP per cycle peak against the
${\sim}2$ of a loop-carried multiply--accumulate datapath, with no
loop-carried arithmetic anywhere in the design.

\item \textbf{EPIC-style VLIW control for a statically schedulable
workload} (Sec.~\ref{sec:arch}): a 4-slot word whose \code{wait\_mask}
dependence bits are honored by a scoreboard, dumb, fast hardware under a
smart compiler, with exhaustive pre-issue validation, fault atomicity, and
a fault-freeze/\code{CAUSE}/IRQ recovery discipline.
Measured on the double-buffering regression program: GEMM busy $9{,}880$ and
DMA busy $2{,}114$ cycles against $10{,}543$ wall cycles, a busy sum of
$11{,}994 > 10{,}543$ that \emph{proves} compute/DMA overlap, at 93.7\%
GEMM-slot occupancy.

\item \textbf{A derived tolerance-verification methodology}
(Sec.~\ref{sec:numerics}, \ref{sec:verif}): an FP64 golden model; a
per-element bound derived from rounding-error analysis of tree reductions
(base $2^{-20}$ for FP32 accumulations, $2^{-7}$ for BF16-stored results,
each scaled by $1+\log_2(K)/8$); exact-equality carve-outs; and evidence of
tightness. A 17-test full-UVM environment runs green with zero scoreboard
failures, the training acceptance test alone sweeps $107{,}108$ elements
($28{,}816$ exact, $78{,}292$ tolerance), and embeds a \emph{semantic}
oracle: an in-testbench
diffusion-MLP training run whose loss must decrease, and does, 56.4 to 26.0
over 20 optimizer steps, plus a transformer attention block verified
forward and backward. The bug catalog is three defects: one RTL (a
sticky-level versus registered-clear race in DMA write-error recovery), one
golden-model, one methodological.

\item \textbf{An honest performance framing} (Sec.~\ref{sec:results}):
per-cycle FLOP accounting measured from the RTL configuration, an
iso-node GPU comparison presented as an explicitly analytical
utilization$\times$energy decomposition ($6$--$10\times$) with every
assumption stated, and a clear separation between measured quantities
(cycle counts, occupancy, element checks, and the sky130
synthesis/post-route measurements of Sec.~\ref{sec:results:synth}) and
analytical extrapolation (scale-out configurations, energy ratios).
\end{itemize}

\section{Background and Rationale}\label{sec:background}

\subsection{The training step DTX executes}

DTX targets the training steps of the two model families that dominate
modern generative practice: diffusion denoisers~\cite{ho2020ddpm,
peebles2023dit} and transformer blocks~\cite{vaswani2017attention}. A
DDPM-style $\varepsilon$-prediction step on a clean sample $x_0$ is
\begin{align}
t \sim \mathcal{U}\{1..T\},&\qquad \varepsilon \sim \mathcal{N}(0,I),\nonumber\\
x_t = a_t\,x_0 + b_t\,\varepsilon,&\qquad
\hat\varepsilon = f_\theta(x_t, t),\nonumber\\
L = \tfrac{1}{N}\lVert \varepsilon-\hat\varepsilon\rVert^2,&\qquad
g = \tfrac{2}{N}(\hat\varepsilon-\varepsilon),
\label{eq:trainstep}
\end{align}
followed by backpropagation of $g$ through $f_\theta$ and an
AdamW~\cite{kingma2015adam,loshchilov2019adamw} update of every weight. The
$T$-step sampler is absent from training; the noising front-end is the single
affine op $a_t x_0 + b_t\varepsilon$ (the vector unit's \code{NOISE} op, fed
by the on-chip Philox Gaussian source). When $f_\theta$ is a transformer
block, the forward graph is
LN${\to}QK^{\mathsf T}{\to}$softmax${\to}{\cdot}V{\to}$proj${\to}$residual${\to}$MLP,
and every linear layer $Y = XW^{\mathsf T}$ spawns two backward products,
$dX = dY\,W$ and $dW \mathrel{+}= dY^{\mathsf T}X$, hence the three GEMM
dataflows, one of which accumulates in FP32 onto its output tile across
micro-batch chunks. Everything else in the step, normalization statistics
and their backward, softmax and its backward, activation derivatives, the
loss gradient, casts, and the elementwise optimizer recurrence over
persistent $(w,m,v)$ state, is a streaming map or a row reduction, which is
exactly the vector unit's op alphabet. Storage is BF16 with FP32
accumulation, the established mixed-precision training
split~\cite{micikevicius2018mixed,kalamkar2019bf16}.

\subsection{Why a systolic array under a VLIW word}

The defining property of this workload is that it is \emph{statically
schedulable}. A training step is a compile-time-known DAG of tile-granular
operations: its shapes are fixed by the model, it contains no data-dependent
control flow, and in DTX every compute engine is a free-running pipeline with
a published latency contract (the \code{LAT\_*} local parameters of
\code{dtx\_pkg}), no stalls, and initiation interval~1, even the Newton
reciprocal/rsqrt and the piecewise-polynomial transcendentals are fixed-depth
pipelines. A machine for such a workload does not need discovery of
parallelism at run time; it needs the parallelism \emph{stated}. DTX
therefore adopts the EPIC position~\cite{fisher1983vliw,schlansker2000epic}:
a 4-slot VLIW word (GEMM, vector, optimizer/RNG, DMA, one slot per engine
class, so structural hazards are encoded by construction) in which the
compiler schedules dependences explicitly as per-slot \code{wait\_mask}
bits. The hardware honors the mask with a scoreboard over pending-or-busy
engines and issues every slot whose conditions hold in the same cycle; the
one genuinely non-deterministic latency in the system, the AXI DMA, is
absorbed by the mask's drain semantics (``named slots from prior words have
completed'') rather than by cycle counting, and ping-pong tile buffers turn
the DMA into a schedulable slot instead of a cache hierarchy. A one-level
zero-overhead loop covers the tile loop nests that dominate training
programs. Within the GEMM slot, the dense tile products that account for
almost all training FLOPs go to a weight-stationary systolic
array~\cite{jouppi2017tpu}, which buys $2\,\mathrm{SYS\_N}^2$ FLOP per cycle
of the one resource a serialized-reduction machine cannot scale: concurrent,
\emph{unordered} multiply-accumulate.

\subsection{Why tolerance verification is the correct contract here}

FP32 addition is not associative, so the value of a sum is a property of its
evaluation \emph{order}. One could make the order normative, %
sequential, ascending index, which would buy byte-comparability at the cost
of the loop-carried floor of Sec.~\ref{sec:intro}. DTX's trees and systolic
columns evaluate a parallel order instead: still \emph{fixed}, hence still
run-to-run reproducible for a given program, but provably different from the
sequential sum by a data-dependent rounding residue. For such hardware,
there are only two honest verification strategies. One is to
replicate the exact tree topology inside the golden model, restoring
bit-equality, but then the reference is a second implementation of the same
pipeline rather than an independent statement of the mathematics, and the
classic twin-reference failure mode (a constant truncated
identically on both sides, making a bit-exact match certify wrong
arithmetic) shows that agreement between twins is not correctness.
The other is to compare against a \emph{higher-precision} independent
reference under a bound derived from rounding-error analysis. Classical
analysis~\cite{higham1993summation,goldberg1991floating} gives the tree the
better error bound, $\lceil\log_2 K\rceil\, u \sum_i|x_i|$ for a balanced
tree versus $(K{-}1)\,u \sum_i|x_i|$ for the sequential chain, with $u$ the
unit roundoff, so the parallel order is numerically \emph{tighter}, not
looser; what is lost is only comparability, and what a correct scoreboard
owes is a tolerance that scales with $\log_2 K$, not with $K$. DTX adopts
this second strategy: an FP64 (\code{real}) golden model, a per-element bound
whose relative term grows as $1+\log_2(K)/8$ from a base of $2^{-20}$ (FP32
accumulations) or $2^{-7}$ (BF16-stored results), an absolute floor of
$10^{-6}$ where flush-to-zero and cancellation make relative error
meaningless, and exact equality wherever the arithmetic is exact, DMA
copies, casts of exact values, the integer Philox stream, and register
readback. Tolerance is not a retreat from rigor; deployed with derived
bounds and exact carve-outs, it is the only contract under which an
unordered-reduction machine can be checked against a reference that is
independent of the device. Sec.~\ref{sec:numerics} derives the bound;
Sec.~\ref{sec:verif} shows it is tight enough to catch a single
out-of-contract rounding hop at $5{,}340\times$ the budget.

\section{Architecture: The Programmer's View}\label{sec:arch}

\begin{figure}[t]
  \centering
  \includegraphics[width=\columnwidth]{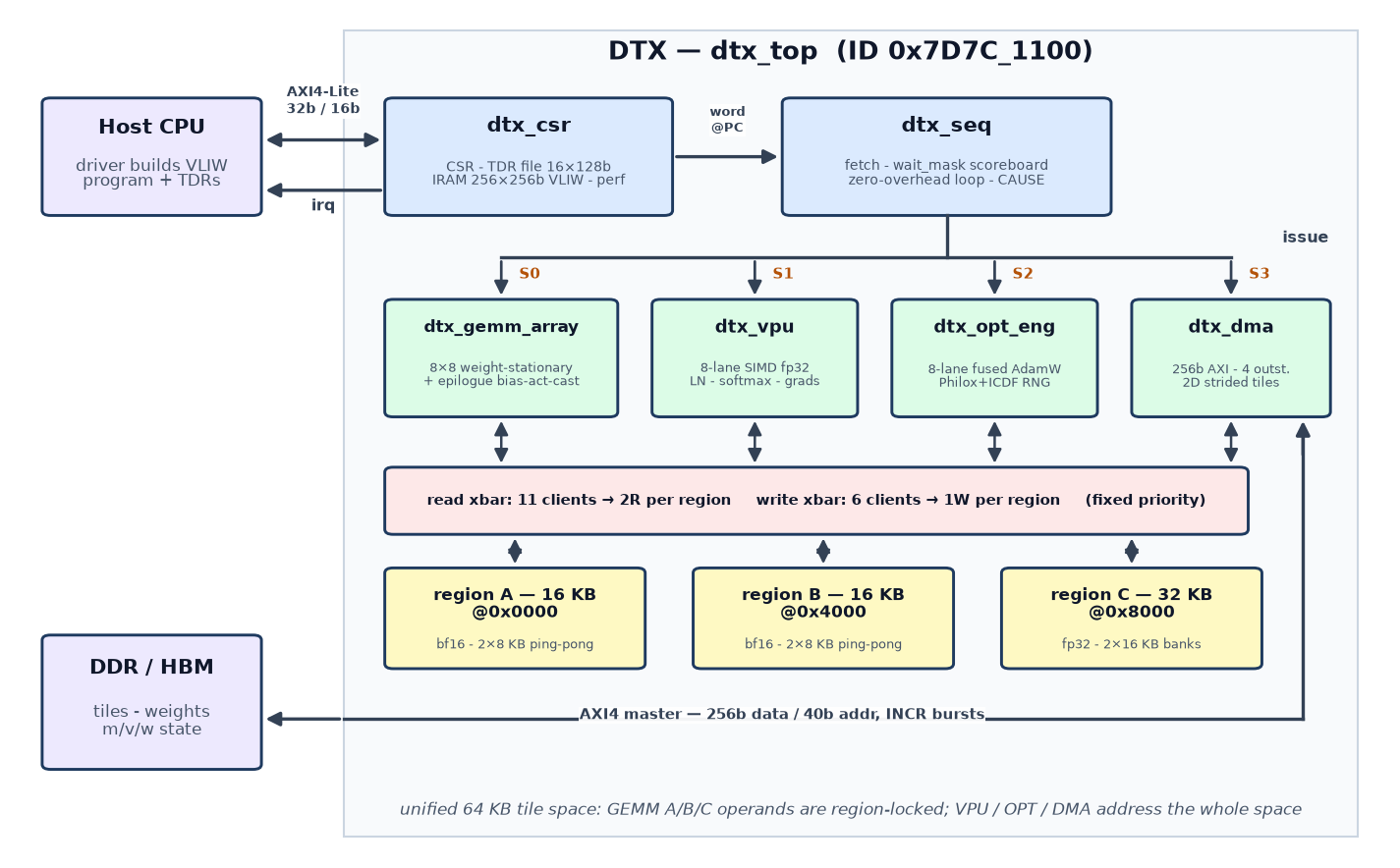}
  \caption{DTX top level. The VLIW sequencer issues four slots onto four
  engines (systolic GEMM + fused epilogue, VPU, OPT/RNG, DMA), which meet in
  a unified 64\,KB tile space of three region memories, each granting two
  reads and one write per cycle.}
  \label{fig:top}
\end{figure}

DTX (\code{ID = 0x7D7C\_1100}) presents to the host as a stored-program VLIW
device behind three wires: one AXI4 master (256-bit data, 40-bit address) for
tensor traffic, one AXI4-Lite slave for the CSR file, the tile-descriptor
window, and the instruction RAM, and one \code{irq}
(Fig.~\ref{fig:top}). The host compiles a training step into 256-bit VLIW
words in a 256-deep instruction RAM, points the operand fields of each slot
at entries of a 16-deep tile-descriptor register (TDR) file, programs the
optimizer and RNG state through the CSR, pulses \code{START}, and polls
\code{DONE|ERR}. Rather than being driven by the host one macro-op
descriptor at a time, DTX exposes an on-chip memory hierarchy: all compute operands live in a unified 64\,KB tile
space that the DMA slot fills and drains under explicit program control.
The contract throughout is \emph{dumb fast hardware, smart compiler}: the
sequencer never reorders, renames, or interlocks beyond what the program
states; every dependence, every double-buffer rotation, and every memory-port
budget is scheduled by the code generator and merely \emph{honored} by the
machine.

\subsection{The 256-bit VLIW Word}\label{sec:arch:word}

\begin{figure}[t]
  \centering
  \includegraphics[width=\columnwidth]{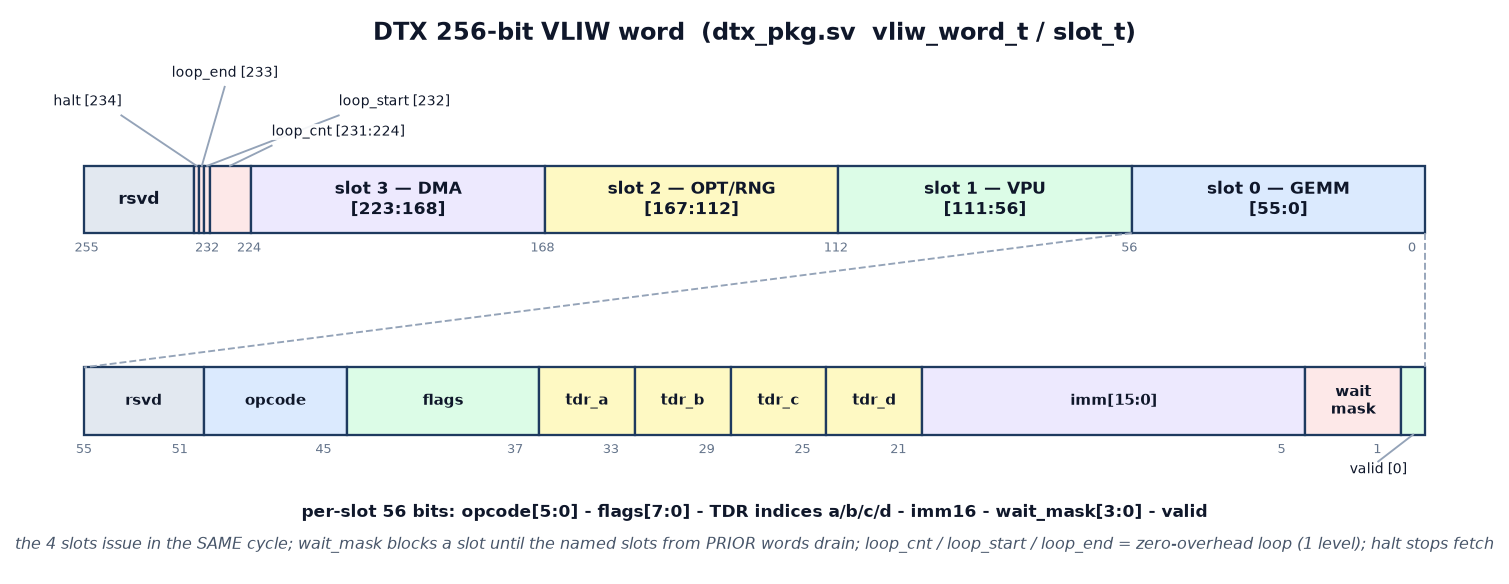}
  \caption{The 256-bit VLIW word: four 56-bit slots (GEMM, VPU, OPT/RNG,
  DMA), each carrying \code{opcode}, \code{flags}, four TDR indices, a 16-bit
  immediate, and a 4-bit \code{wait\_mask}; word-level bits control the
  single-level zero-overhead loop and \code{halt}.}
  \label{fig:vliw}
\end{figure}

Figure~\ref{fig:vliw} shows the instruction format (\code{dtx\_pkg.sv}
\code{slot\_t}/\code{vliw\_word\_t}). A word is four 56-bit slots plus a
word-level control byte. Slot $k$ is statically bound to engine $k$: slot~0
GEMM (bits [55:0]), slot~1 VPU ([111:56]), slot~2 OPT/RNG ([167:112]),
slot~3 DMA ([223:168]). Within a slot, LSB-first:
\code{valid} [0]; \code{wait\_mask} [4:1], one bit per peer slot;
\code{imm} [20:5], a 16-bit op-specific immediate; four 4-bit TDR indices
\code{tdr\_d} [24:21], \code{tdr\_c} [28:25], \code{tdr\_b} [32:29],
\code{tdr\_a} [36:33]; an op-specific \code{flags} byte [44:37]; a 6-bit
per-slot \code{opcode} [50:45]; and reserved bits [55:51]. The word-level
fields are \code{loop\_cnt} [231:224], \code{loop\_start} [232],
\code{loop\_end} [233], and \code{halt} [234]; bits [255:235] are reserved.
Operands are named \emph{indirectly}: a slot carries only 4-bit TDR indices,
and the 128-bit descriptors live in an architectural register file, this
indirection is what keeps four fully-specified tensor operations inside
256 bits.

Each slot has its own opcode space. Slot~0: \code{G\_FWD}
($Y{=}XW^{\mathsf T}$ with fused epilogue), \code{G\_BWD\_DX}
($dX{=}dY\,W$), and \code{G\_BWD\_DW} ($dW{\mathrel{+}=}dY^{\mathsf T}X$,
FP32-accumulating onto the C operand); the GEMM \code{flags} byte selects the
epilogue, activation (\code{NONE}/\code{RELU}/\code{GELU}) in bits [1:0],
\code{+bias} row (operand D), accumulate-onto-C, and cast-to-BF16-on-drain.
Slot~1 is the SIMD vector unit: LayerNorm forward/backward, softmax
forward/backward, activation backward, elementwise add/multiply,
BF16$\leftrightarrow$FP32 cast, column-sum bias backward, MSE and
cross-entropy gradients, and the closed-form noising op
$x_t{=}a_t x_0{+}b_t\varepsilon$. Slot~2 runs the fused AdamW step
(\code{O\_ADAMW}) or streams Philox-derived randomness into a tile
(\code{O\_RNG\_GAUSS}/\code{O\_RNG\_UNIF}). Slot~3 is the DMA:
2-D strided tile gather/scatter (\code{D\_LD\_TILE}/\code{D\_ST\_TILE}) and
descriptor load (\code{D\_LDTDR}). Opcode 0 in every space is \code{NOP};
a slot with \code{valid}=0 or a \code{NOP} opcode is pre-retired at decode
and never touches its engine.

\subsection{Co-Issue and the \code{wait\_mask} Dependence Discipline}
\label{sec:arch:waitmask}

All four slots of a word issue \emph{concurrently}. The sequencer
(\code{dtx\_seq}) fetches a word, validates all of it
(Sec.~\ref{sec:arch:err}), and then fires each valid slot independently the
cycle both of two conditions hold: (a)~no engine named in the slot's
\code{wait\_mask} is pending or busy, and (b)~the slot's own engine is free
(the structural hazard is implicit; back-to-back ops in the same slot never
need a mask bit). Slots whose conditions hold in the same cycle issue in the
same cycle, the EPIC common case, and a blocked slot delays only itself and
the advance to the next word, never a peer that already fired.

\code{wait\_mask} semantics are deliberately minimal: bit $k$ means ``do not
issue until the engine of slot $k$, as occupied by \emph{prior} words, has
drained.'' There is no register scoreboard, no operand comparison, and no
hardware discovery of dependences: RAW, WAR, and WAW hazards through the tile
space, and, equally important, \emph{memory-port budget} conflicts
(Sec.~\ref{sec:arch:tile}), are all resolved at compile time by placing mask
bits, in the explicit-dependence tradition of EPIC and the TI~C6x VLIW DSPs.
The hardware honors the mask against a per-engine \code{pend|busy}
scoreboard: \code{pend} is set on the issue cycle and cleared by the engine's
\code{done} pulse, so the one-cycle skew between \code{req.valid} and the
engine raising \code{busy} is architecturally invisible. A mask bit naming a
slot of the \emph{same} word is well defined but issue-order dependent; the
compiler contract forbids it. One visibility rule follows from the pipeline:
TDR indices are resolved when a word is \emph{checked}, so a program that
rewrites a descriptor with \code{D\_LDTDR} must set the DMA bit in the
consuming word's \code{wait\_mask}, same-word slots always see the old
descriptor.

\subsection{Tile Descriptor Registers}\label{sec:arch:tdr}

The TDR file is 16 architectural registers of 128 bits, written either
through the CSR window (offsets \code{0x100}--\code{0x1FF}, four 32-bit
subwords each) or from memory by \code{D\_LDTDR}. The resolved fields,
LSB-first, are \code{base}[39:0] (byte address), \code{pitch}[55:40] (bytes
between rows), \code{rows}[63:56], \code{cols}[71:64] (each 1--64), and
\code{fmt}[73:72] (\code{BF16} or \code{FP32}); bits [127:74] are reserved.
One descriptor shape serves two address spaces: for a DMA memory operand,
\code{base} is a 40-bit external AXI byte address; for every compute operand
(and the DMA's tile side, Sec.~\ref{sec:arch:tile}), it is a byte address in
the 64\,KB unified tile space. The file is shared between both uses, and a
collision, one index doing double duty as a tile and a memory
descriptor, silently retargets an operand rather than faulting; the software
convention is therefore tiles in TDR\,0--7 and memory descriptors in
TDR\,8--15, and the shipped programs (Listing~\ref{lst:trainstep}) follow it.

\subsection{Zero-Overhead Loops}\label{sec:arch:loop}

DTX supports one hardware loop level. A word with \code{loop\_start} set and
\code{loop\_cnt}${}\neq 0$ opens a loop whose body executes \code{loop\_cnt}
times (8-bit count, up to 255 executions); the word with \code{loop\_end}
takes the back-edge to the opening word while iterations remain. The redirect
happens in the sequencer's normal PC-advance cycle, so the back-edge costs
zero additional cycles, exactly what a $\lceil K/8\rceil{\times}\lceil
N/8\rceil$-pass tiled GEMM schedule or a $T$-step sampling sweep wants.
\code{loop\_start} on the re-entered word is ignored while the loop is
active (required for re-entry); nesting is a compiler-contract violation, and
\code{halt} has priority over a pending back-edge.

\subsection{The Unified 64\,KB Tile Space and Its Region Budget}
\label{sec:arch:tile}

All engine operands live in one 64\,KB byte-addressed tile space, physically
three region memories (\code{dtx\_tbmem}) of 256-bit words: region~A at
\code{0x0000} (16\,KB, two 8\,KB ping-pong banks selected by
\code{base}[13]), region~B at \code{0x4000} (16\,KB, likewise), and region~C
at \code{0x8000} (32\,KB, two 16\,KB banks selected by \code{base}[14]).
A and B hold BF16 tiles, C holds FP32 tiles. Each region grants \emph{two
reads and one write per cycle} through a fixed-priority crossbar serving 11
read clients (GEMM A/B/C, DMA, VPU A/B/C, OPT $m$/$v$/$g$/$w$) and 6 write
clients (epilogue-C, DMA, VPU, OPT $m'$/$v'$/$w'$-or-RNG). A request beyond
the 2R+1W budget is \emph{dropped}, a program error, \code{\$error}-ed in
simulation and deterministic garbage in hardware, because the budget, like
the hazards, is a compile-time obligation discharged with \code{wait\_mask}
scheduling. Reads land one cycle after grant and every engine consumes read
data on the landing cycle, so port sharing never corrupts a granted access.

Three placement rules make up the program contract. First, \emph{GEMM region
binding}: the array's ports are hard-wired to the regions, so the A operand
must live in region~A, the B operand and the bias row (operand D) in
region~B, and the C tile in region~C; bases must be SYS\_N-element aligned
and in the correct region, else \code{ERR\_BAD\_SIZE}. While a GEMM runs it
occupies one read port on all three regions, plus the epilogue's C write port
during drains. Second, the \emph{AdamW three-region rule}: \code{O\_ADAMW}
streams four reads ($m,v,g,w$) and three writes ($m',v',w'$) concurrently, so
$m$, $v$, and $w$ must occupy three \emph{different} regions and only $g$ may
share with one of them, canonically $m{@}A$, $v{@}B$, $w{@}C$, $g{@}C$; the
update is in-place ($m'{\to}a$, $v'{\to}b$, $w'{\to}d$), race-free because
writes trail reads by the fixed pipe latency. Third, an affine LayerNorm
reads $\gamma$ and $\beta$ through two ports of one region, which therefore
must not also host the LN input tile.

The DMA addresses the same space through its immediate: for
\code{D\_LD\_TILE}/\code{D\_ST\_TILE}, \code{imm}[10:0] is the destination
(source) tile-space \emph{word} address and \code{imm}[15:11] is the
tile-side row pitch in 256-bit words, with 0 meaning packed. Memory row $r$
is fetched from \code{base}${}+r\cdot{}$\code{pitch} (any byte alignment,
$\mathit{cols}\times\mathit{esize}$ bytes) and packed into
$\lceil\mathit{cols}\cdot\mathit{esize}/32\rceil$ consecutive tile words at
$\code{imm}+r\cdot\mathit{tb\_pitch}$, tail words byte-enabled;
$\mathit{tb\_pitch}$ smaller than the packed row width is
\code{ERR\_BAD\_SIZE}. GEMM-consumable tiles use the fixed-pitch convention
the array expects: pitch code 4 for BF16 A/B rows (128\,B) and 8 for FP32 C
rows (256\,B). Stores mirror the layout exactly, masking partial beats with
\code{WSTRB}.

\subsection{CSR Map}\label{sec:arch:csr}

\begin{table}[t]
  \centering
  \caption{AXI4-Lite CSR map (byte offsets). OPT hyperparameters are full
  32-bit FP32 bit patterns, quasi-static (write only while \code{!BUSY}).}
  \label{tab:csr}
  \footnotesize
  \begin{tabularx}{\columnwidth}{@{}l l X@{}}
    \toprule
    Offset & Register & Function \\
    \midrule
    \code{0x000} & \code{ID} & RO, \code{0x7D7C\_1100} \\
    \code{0x004} & \code{CTRL} & W1P: [0] \code{START}, [1] \code{ABORT},
      [8] \code{CNT\_CLR} \\
    \code{0x008} & \code{STATUS} & [0] \code{BUSY}, [1] \code{DONE},
      [2] \code{ERR}, [3] \code{HALTED}, [15:8] PC \\
    \code{0x00C/10} & \code{IRQ\_EN/STAT} & [0] done, [1] err; \code{STAT}
      is W1C \\
    \code{0x014} & \code{PC} & write start PC; read current/start \\
    \code{0x018} & \code{CAUSE} & RO fault record \{pc, slot, err\} \\
    \code{0x01C--24} & \code{SEED0/1, STREAM} & Philox key and stream id \\
    \code{0x028--50} & \code{PERF\_*} & \code{CYCLES}; per-slot
      \code{BUSY0..3}, \code{STALL0..3}; DMA bus bytes rd/wr \\
    \code{0x054--6C} & \code{OPT\_*} & \code{LR}, \code{BETA1}, \code{BETA2},
      \code{EPS}, \code{LRWD}, \code{RB1K}, \code{RB2K} (FP32) \\
    \code{0x070} & \code{OPT\_STAT} & [0] sticky \code{nan\_seen}, W1C \\
    \code{0x074} & \code{RNG\_CTR} & Philox block counter; live readback \\
    \code{0x100--1FF} & TDR window & $16\times128$\,b, $4\times32$-bit
      subwords \\
    \code{0x1000--2FFF} & IRAM window & $256\times256$\,b, $8\times32$-bit
      subwords \\
    \bottomrule
  \end{tabularx}
\end{table}

Table~\ref{tab:csr} summarizes the AXI4-Lite map. Control semantics are
deliberately conventional: \code{CTRL} is write-one-to-pulse, \code{START} is
accepted only when idle and clears \code{DONE}/\code{ERR}/\code{HALTED} and
both IRQ bits, \code{ABORT} is accepted only while busy and wins over a
same-write \code{START}. \code{STATUS.HALTED} is the \code{DONE|ERR} level;
drivers poll \code{DONE|ERR}. Performance counters expose exactly the
quantities the co-issue architecture makes interesting: total busy cycles,
per-slot busy and per-slot issue-stall levels, and true DMA bus bytes in each
direction, these counters are how the compute/DMA overlap proof of
Sec.~\ref{sec:results} is measured, from software, on the shipped device.

The optimizer hyperparameters deserve emphasis. \code{OPT\_LR},
\code{OPT\_BETA1/2}, \code{OPT\_EPS} (added \emph{outside} the square root,
the PyTorch convention), \code{OPT\_LRWD} (the host-computed product
$\eta\lambda$), and the bias corrections \code{OPT\_RB1K/RB2K}
($1/(1-\beta_i^k)$) are all \emph{full 32-bit FP32 bit patterns} in the CSR,
never values squeezed through the 16-bit slot immediate. This designs out a
subtle and dangerous defect class: a
bias-correction constant truncated identically by RTL and golden model
produces an exact match on wrong arithmetic, a defect that surfaces only
when a real model fails to train. In DTX no optimizer constant ever passes
through a narrowing field, and the end-to-end training test of
Sec.~\ref{sec:verif} keeps the semantic oracle in the regression.

\subsection{Error Model and Recovery}\label{sec:arch:err}

Every fault resolves to one of six codes: \code{ERR\_ABORT},
\code{ERR\_BAD\_OP}, \code{ERR\_BAD\_SIZE}, \code{ERR\_AXI\_RD},
\code{ERR\_AXI\_WR}, \code{ERR\_BAD\_PC}. Validation is exhaustive and
happens \emph{before} issue, in two phases. The static phase checks, per
slot in parallel: opcode within the slot's legal space and flag encodings
well-formed (the reserved activation encoding \code{2'b11} rejects
\code{ERR\_BAD\_OP}); descriptor dimensions in 1--64, format legal, and
base/pitch element-aligned (\code{D\_LDTDR} sources must be 16-byte aligned),
else \code{ERR\_BAD\_SIZE}. The footprint phase then evaluates, one operand
column per cycle across all four slots, the 48-bit inequality
$\code{base} + (\mathit{rows}{-}1)\cdot\mathit{pitch} + \mathit{row\_bytes}
\le 2^{40}$ for every operand an opcode architecturally uses. The whole word
is validated before \emph{any} slot of it issues, fault atomicity: a word
never partially executes because a later slot was malformed. A PC running
past the 256-word IRAM is \code{ERR\_BAD\_PC}.

On any fault, static, footprint, a runtime engine error (a \code{done}
pulse carrying \code{err}${}\neq{}$\code{ERR\_NONE}, e.g.\ an AXI
\code{RRESP}/\code{BRESP} error surfacing as
\code{ERR\_AXI\_RD}/\code{WR}), or a host abort (reported as
\code{ERR\_ABORT}), the sequencer freezes fetch, latches
\code{CAUSE}~=~\{\code{pc}[7:0], \code{slot}[1:0], \code{err}[3:0]\}, raises
the error interrupt, and asserts a level abort to all engines, which quiesce
without emitting \code{done}; the sequencer force-clears its scoreboard,
waits for the DMA burst engines to drain (AXI protocol is never violated
mid-burst; an aborted store completes its issued bursts with
\code{WSTRB}=0), and returns to idle with \code{STATUS.ERR} set. Runtime
errors are attributed to the \emph{issuing} word: the sequencer records the
PC each engine is executing for, so \code{CAUSE.pc} names the faulting
instruction even when the error arrives many cycles after issue, with other
slots in flight. First fault wins; subsequent errors during the drain do not
overwrite \code{CAUSE}. Recovery is entirely host-driven and needs no reset:
poll \code{DONE|ERR}, read \code{CAUSE}, clear \code{IRQ\_STAT} (W1C), fix or
skip the offending program, and pulse \code{START}, which is exactly the
fault-then-clean-rerun sequence the verification suite drives
(Sec.~\ref{sec:verif}), and the path on which the one RTL bug of this project
was found.

\subsection{A Diffusion Training Step as a VLIW Program}
\label{sec:arch:program}

Listing~\ref{lst:trainstep} shows the training step the regression actually
runs (condensed from \code{dtx\_train\_mlp\_seq.sv}): one iteration of a
denoiser-MLP diffusion step, noise, forward, MSE gradient, backward, fused
AdamW, expressed as five short wait-mask-chained VLIW programs that the host
launches in turn, staging operands between them (the staging rationale is a
verification-methodology story told in Sec.~\ref{sec:verif}). Each program is
self-contained: loads, compute, stores, \code{halt}.

\begin{lstlisting}[language=SV, float=t,
  caption={One iteration of the diffusion training step as VLIW programs
  (condensed from \code{dtx\_train\_mlp\_seq.sv}). \code{T0--T7} are tile
  TDRs, \code{T8--T15} memory TDRs; \code{wait} sets name slots
  \{G,V,O,D\}.},
  label={lst:trainstep}]
// P_noise: eps = RNG; x_t = a_t*x0 + b_t*eps
w0: DMA LD_TILE  T8 ->A0 (x0)            wait {}
w1: DMA LD_TILE  T9 ->C0 (a_t,b_t)       wait {D}
w2: OPT RNG_GAUSS c=T1 (eps->A1)         wait {}
w3: VPU NOISE  a=T0 b=T1 c=T2 d=T3       wait {O,D}
w4: DMA ST_TILE  T10<-A1 (eps)           wait {V}
w5: DMA ST_TILE  T11<-A2 (x_t)           wait {D} halt

// P_fwd: Y=relu(x_t*W'+b); g=Y-eps; dpre=g.relu'(Y)
w0..w3: LD x_t->A0, W->B0, bias->B1, eps->A1
w4: GEMM FWD   a=T0 b=T1 c=T4 d=T2       wait {D}
               flags = BIAS|RELU         // Y -> C0
w5: VPU MSE_GRAD a=T4(Y) b=T3(eps) d=T5  wait {G}
w6: VPU ACT_BWD  a=T5 b=T4 d=T6 (relu')  wait {V}
w7: DMA ST_TILE  T12<-C1 (g)             wait {V}
w8: DMA ST_TILE  T13<-C2 (dpre)          wait {V} halt

// P_grad: dW=dpre'*x_t; dX=dpre*W; db=colsum(dpre)
w0..w4: LD dpre'->A0, x_t->B0, dpre->A1,
           W->B1, 0->C0 (dW accum preload)
w5: GEMM BWD_DW  a=T0 b=T1 c=T4 (dW+=)   wait {D}
w6: GEMM BWD_DX  a=T2 b=T3 c=T5          wait {G}
w7: VPU BIAS_BWD a=T2 d=T6 (db)          wait {G}
w8: DMA ST_TILE  T13<-C0 (dW)            wait {G}
w9: DMA ST_TILE  T14<-C2 (db)            wait {V} halt

// P_optW: AdamW -- 3-region rule m@A v@B w@C g@C
w0..w3: LD m->A0, v->B0, g->C0, w->C1
w4: OPT ADAMW  m=T0 v=T1 g=T2 w=T3       wait {D}
               flags = DECOUPLED_WD      // in place
w5: DMA ST_TILE  T12<-A0 (m')            wait {O}
w6: DMA ST_TILE  T13<-B0 (v')            wait {O}
w7: DMA ST_TILE  T14<-C1 (w')            wait {O} halt
// P_optB: same shape on the bias vector
\end{lstlisting}

The listing exercises most of the contract in nine words at a time. In
\code{P\_noise}, the RNG op at \code{w2} carries an empty mask and therefore
\emph{runs concurrently} with the still-draining coefficient load from
\code{w1}, engine-level parallelism falls out of simply not naming a
dependence, while \code{NOISE} waits on both producers with
$\{O,D\}$. In \code{P\_fwd}, one GEMM word encodes the entire fused forward
layer ($\times W^{\mathsf T}$, $+b$, ReLU) via the epilogue flags, and the
two VPU words chain RAW hazards down the mask bits
$\{G\}{\to}\{V\}{\to}\{V\}$. In \code{P\_grad}, the mask on \code{w7} is a
\emph{port-budget} wait, not a data dependence: \code{BIAS\_BWD} writes
region~C, which the epilogue's write port owns until the \code{BWD\_DX}
drain completes. \code{P\_optW} shows the three-region AdamW placement and
the in-place update, with the three result stores fanned out after a single
$\{O\}$ wait. Run for 20 iterations in the testbench with fixed noise, this
program's loss falls from 56.4 to 26.0 (Sec.~\ref{sec:results}), the ISA's
end-to-end semantic check, not merely a per-op one.

\section{Numerics: Tolerance as the Contract}\label{sec:numerics}

DTX's numerics are not a reduction-order contract. The normative content
is instead (i) a structural law on how FP
hardware may be built, (ii) a set of unit-level primitives that remain
bit-exact against their own reference functions, and (iii) a
\emph{tolerance model}, a derived, per-element error bound against an
FP64 golden model, that replaces byte-equality wherever throughput
hardware reorders arithmetic.

\subsection{The pipelined-FP law}

Combinational depth is what kills large FP designs at technology
mapping: a ten-round combinational Philox, for instance, is eighty
$32{\times}32$ multiplies per beat, beyond what ABC can map at chip
scale. DTX forecloses the failure mode with a \emph{design law}, stated in
the RTL and honored by every module: every FP operation is a pipelined unit (\code{fp32}
add/mul and \code{bf16} mul are 2-stage); no combinational chain is deeper
than one FP op; Newton loops are unrolled, one iteration per pipe segment;
there are no data-dependent \code{while} loops; and \emph{loop-carried
accumulation is a design bug}, reductions are trees or latency-interleaved
banks (Sec.~\ref{sec:uarch}). The corollary is the free-running discipline:
every unit is II${=}1$ with a fixed \code{in\_valid}$\to$\code{out\_valid}
latency, published as the \code{dtx\_pkg} \code{LAT\_*} timing contract
(Table~\ref{tab:fpunits}), so composition is delay-line matching, never
handshaking.

\begin{table}[t]
  \caption{FP primitive timing contract (\code{dtx\_pkg} \code{LAT\_*})
  and accuracy. ``$\equiv$ pkg'' units carry simulation shadow assertions
  checking every emitted result, bit for bit, against the package
  reference function the golden model also calls. PWL bounds are
  exhaustive over all input codes; contract bounds from the architecture
  plan in parentheses.}
  \label{tab:fpunits}
  \centering
  \footnotesize
  \begin{tabular}{@{}lccl@{}}
    \toprule
    Unit & Lat. & II & Accuracy \\
    \midrule
    \code{dtx\_bf16\_mul}  & 2  & 1 & exact (no rounding exists) \\
    \code{dtx\_fp32\_add/mul} & 2 & 1 & RNE, FTZ; $\equiv$ pkg \\
    \code{dtx\_recip\_nr}  & 19 & 1 & $\equiv$ pkg (3 NR iterations) \\
    \code{dtx\_rsqrt\_nr}  & 25 & 1 & $\equiv$ pkg (3 NR iterations) \\
    \code{dtx\_exp2\_pwl}  & 4  & 1 & rel $\le 1.12{\times}10^{-5}$
                                      ($10^{-3}$) \\
    \code{dtx\_gelu\_pwl}  & 5  & 1 & abs $\le 2.21{\times}10^{-4}$
                                      ($2{\times}10^{-3}$) \\
    \code{dtx\_gelu\_dpwl} & 5  & 1 & abs $\le 8.0{\times}10^{-4}$
                                      ($2{\times}10^{-3}$) \\
    \code{dtx\_philox\_pipe} & 10 & 1 & $\equiv$ pkg (bit-exact) \\
    \bottomrule
  \end{tabular}
\end{table}

\subsection{Formats and exact primitives}

Storage is BF16, accumulation and all internal arithmetic
FP32~\cite{kalamkar2019bf16}; rounding is round-to-nearest-even
everywhere; subnormals flush to zero on every unit's inputs and outputs.
The BF16$\times$BF16 product is \emph{exact} in FP32 (the $8{\times}8$
significand product occupies at most 16 bits of the 24-bit mantissa;
guard and sticky are structurally zero, so RNE is a no-op), and
\code{cast\_bf16} is the single narrowing routine. \code{dtx\_fp32\_add}
folds specials, magnitude swap, and the alignment shift (sticky collapsed
into bit~0) into stage~1, and add/subtract, leading-zero-count
normalization, RNE, and clamps into stage~2; \code{dtx\_fp32\_mul} is the
analogous two stages around a registered $24{\times}24$ product. Both are
bit-identical to the \code{dtx\_pkg} functions the golden model imports,
and that identity is not assumed but
continuously \emph{checked}: each instance carries an
\code{ifndef SYNTHESIS} shadow queue asserting every output against the
reference function, so hundreds of instances re-prove the primitive on
every regression cycle they compute.

\subsection{Magic-constant Newton, one iteration per segment}

$1/x$ and $1/\sqrt{x}$ (LayerNorm, softmax, AdamW) are fixed-count
Newton--Raphson from integer magic seeds, \code{0x7EF127EA} for the
reciprocal, \code{0x5F3759DF}~\cite{lomont2003invsqrt} for
rsqrt, with $\code{NR\_ITERS}{=}3$ iterations \emph{unrolled into the
pipelined units}: an iteration of $y' = y(2-xy)$ is
mul$\,{+}\,$sub$\,{+}\,$mul $=6$ cycles, giving $\code{LAT\_RECIP} = 1 +
3(2\cdot2+2) = 19$; an iteration of $y' = y(1.5 - \tfrac{x}{2}y^2)$ is
$8$ cycles, giving $\code{LAT\_RSQRT} = 1 + 3(3\cdot2+2) = 25$. The seed's
${\sim}5{\times}10^{-2}$ relative error squares per iteration to
${\sim}4{\times}10^{-11}$, below one FP32 ulp. Specials ($1/0$,
$\mathrm{rsqrt}(\pm0)$, negatives, NaN/Inf) are decoded at stage~0 and ride
a latency-matched bypass; $x/2$ is an exponent decrement, bit-identical to
$\code{fp32\_mul}(0.5,x)$ on the normal path. Both units shadow-assert
bit-identity to \code{recip\_f32}/\code{rsqrt\_f32}, so the AdamW
denominator equals \code{adam\_rdenom} exactly, NaN cases included.

\subsection{Piecewise-quadratic transcendentals}

$2^x$ (softmax) and GELU (forward and derivative) are 16-segment,
degree-2 piecewise polynomials evaluated in \emph{pure fixed point}, no
FP ops inside, one integer multiply per stage, so the law above is
trivially honored. \code{dtx\_exp2\_pwl} (4-stage) converts FP32 to s8.16
fixed point ($n = \lfloor x\rfloor$, 16-bit fraction), selects the segment
on the fraction's top nibble, runs two Horner steps against u.24
coefficient ROMs (quadratics through segment endpoints and midpoint of
$2^f$), and reassembles with exponent $127{+}n$; out-of-range inputs
saturate to $+\infty$ / FTZ $+0$ and $|x| < 2^{-17}$ returns exactly
$1.0$. Exhaustively over all $2^{16}$ fraction codes, the PWL core errs
$\le 7.26{\times}10^{-7}$ relative and input truncation adds
$\le 1.06{\times}10^{-5}$; the measured total is $1.12{\times}10^{-5}$
against a $10^{-3}$ contract. \code{dtx\_gelu\_pwl} (5-stage) uses the
same geometry on $|x|\in[0,4)$ ($h{=}0.25$, s3.22 coefficients fit to the
exact erf-form GELU), closing the negative half by the identity
$\mathrm{gelu}(-a) = \mathrm{gelu}(a) - a$ and the tail by
$\mathrm{gelu}(x)=x$ ($x\ge4$) $/$ $-0$ ($x\le-4$): core
$9.7{\times}10^{-5}$, truncation $1.33{\times}10^{-4}$, tail cutoff
$1.27{\times}10^{-4}$, measured total $2.21{\times}10^{-4}$ absolute.
\code{dtx\_gelu\_dpwl} evaluates $\mathrm{gelu}'(x) = \Phi(x) + x\phi(x)$
for \code{V\_ACT\_BWD} with symmetry $\mathrm{gelu}'(-a) = 1 -
\mathrm{gelu}'(a)$ and tail $1.0$ $/$ $+0$; total $\le 8.0{\times}10^{-4}$
absolute, exhaustive, with $\mathrm{gelu}'(\pm0)=0.5$ exact. These
measured envelopes are not decoration: the golden model injects them as
per-tensor tolerance extras (below), so the scoreboard budget for an op
is exactly its documented PWL error plus rounding.

\subsection{Tree reductions, and what they give up}

\code{dtx\_tree\_add} is an $N$-input balanced tree of pipelined adders:
latency $2\log_2 N$, one vector per cycle. \code{dtx\_tree\_max} maps FP32
compare onto \emph{unsigned integer} compare via the order-preserving key
$k = \{\lnot s,\ s\,?\,\lnot m : m\}$, no FP arithmetic, one level per
cycle; NaN carries a large key and propagates as the max (the same poison
behavior an arithmetic reduction would give), and $-\infty$ is the padding
identity. GEMM's $K$-accumulation (down-column, chunk-by-chunk through the
C SRAM) and the VPU's group-then-row re-tree are equally fixed orders.
Every DTX reduction is therefore \emph{deterministic run-to-run}, same
program, same bits, but none equals an ascending sequential chain. Such a
chain is precisely the bottleneck of
Sec.~\ref{sec:intro}: a loop-carried FP32 add caps a
machine at 2\,FLOP/cycle and its fmax at the adder's combinational depth.
Giving up cross-order bit-equality buys $2\,\code{SYS\_N}^2$ GEMM
FLOP/cycle and a design with no loop-carried FP anywhere; what must be
re-established is a verification contract that is neither vacuous nor
brittle. That contract follows.

\subsection{The tolerance model}

The scoreboard checks every non-exact element against the FP64 golden
model:
\begin{align*}
|x_{\mathrm{dut}} - x_{\mathrm{ref}}| &\le
  (10^{-6} + a_x) + (\rho_K + r_x)\,|x_{\mathrm{ref}}|,\\
\rho_K &= \rho_0\bigl(1 + \tfrac{\log_2 K}{8}\bigr),\quad
\rho_0 = \begin{cases} 2^{-7} & \text{BF16-stored}\\
                       2^{-20} & \text{FP32}\end{cases}
\end{align*}
with $K$ the per-tensor reduction depth and $a_x/r_x$ per-tensor extras
computed by the golden model; the whole bound scales by a
\code{tol\_scale} environment knob reserved for long chained programs.
Each term is derived, not tuned:

\emph{FP32 base $2^{-20} = 16u$} ($u = 2^{-24}$, FP32 RNE roundoff). A
depth-$d$ tree applies at most $d$ roundings along any addend's path,
giving the classical bound $|\mathrm{fl}(\Sigma) - \Sigma| \le d\,u
\sum|x_i|$, versus $K u$ for a sequential sum, which is precisely why a
tree is verified by tolerance rather than by equality to an
order-specified reference. The depth-independent floor of $16u$ absorbs
the handful of chained non-reduction ops (epilogue add, cast staging) and
mild $\sum|x_i| / |\Sigma|$ amplification.

\emph{Depth term $(1 + \log_2 K/8)$.} Adds $2^{-20}/8 = 2u$ of relative
budget per tree level: a linear-in-depth allowance with a 2-ulp-per-level
slope, i.e.\ $2\times$ margin over the 1-ulp-per-level worst case. At the
maximum tile depth $K{=}64$ the factor is $1.75$, so
$\rho_K \approx 1.67{\times}10^{-6}$.

\emph{BF16 base $2^{-7}$.} A BF16-stored result's error is dominated by
the single final narrowing ($\le 2^{-8}$ relative, half the $2^{-7}$ ulp
spacing); the base is twice that, covering the double rounding of an
already-rounded FP32 value, and the same depth shape keeps the underlying
FP32 tree uniformly covered.

\emph{Absolute floor $10^{-6}$.} Where $|x_{\mathrm{ref}}|\approx 0$,
relative error is meaningless: the DUT flushes subnormals
($<2^{-126}\to0$) and near-total cancellation legitimately leaves
noise-scale residues.

\emph{Per-tensor extras $a_x/r_x$.} The golden model tracks
$\sum|\mathrm{terms}|$ per output element and folds in the documented PWL
envelopes ($1.2{\times}10^{-5}$ rel for \code{exp2}, $2.3{\times}10^{-4}$
and $8{\times}10^{-4}$ abs for GELU and its derivative) plus cancellation
slack, so ops that use approximate units are budgeted their measured
error and nothing more.

\emph{Exact classes.} Tolerance is granted only where reordering exists.
DMA copies, casts of exact values, CSR/TDR/IRAM readback, and Philox
uniforms (an integer datapath; the $u[31{:}8]\cdot2^{-24}$ conversion is
exact) are compared with \code{===}; NaN must meet NaN and infinities must
match exactly.

The bound is tight rather than vacuous: when a test deliberately broke its
premise, feeding a computed BF16 tile into a GEMM whose reference
consumed the exact FP32 values, a $\sim2^{-8}$ input perturbation against
a $\sim2^{-20}$ output budget, the scoreboard measured a $5{,}340\times$
violation (finding B3, Sec.~\ref{sec:verif}), while all $78{,}292$
tolerance-checked elements of the compliant end-to-end training suite pass
with zero failures. Bit-exactness at the unit level, a derived tolerance
at the tensor level: that split, not either extreme, is what lets a
tree-reduction machine be verified as strictly as an ordered one.

\section{Microarchitecture}\label{sec:uarch}

Every DTX datapath obeys one structural template, stated once here and
enforced by construction (Sec.~\ref{sec:numerics}): a \emph{free-running,
fixed-latency pipeline}, \code{in\_valid} produces \code{out\_valid} a
constant number of cycles later, initiation interval~1, no stalls, no
back-pressure. Engines are then thin FSM wrappers that issue at full rate
and \emph{count completions}; correctness never depends on a pipeline being
able to pause. The SIM configuration is $\code{SYS\_N}{=}8$,
$\code{LANES}{=}8$, tiles up to $64{\times}64{\times}64$, 256-bit tile
words; the PD configuration halves both to $4$.

\subsection{GEMM: weight-stationary systolic array}

\begin{figure}[t]
  \centering
  \includegraphics[width=\columnwidth]{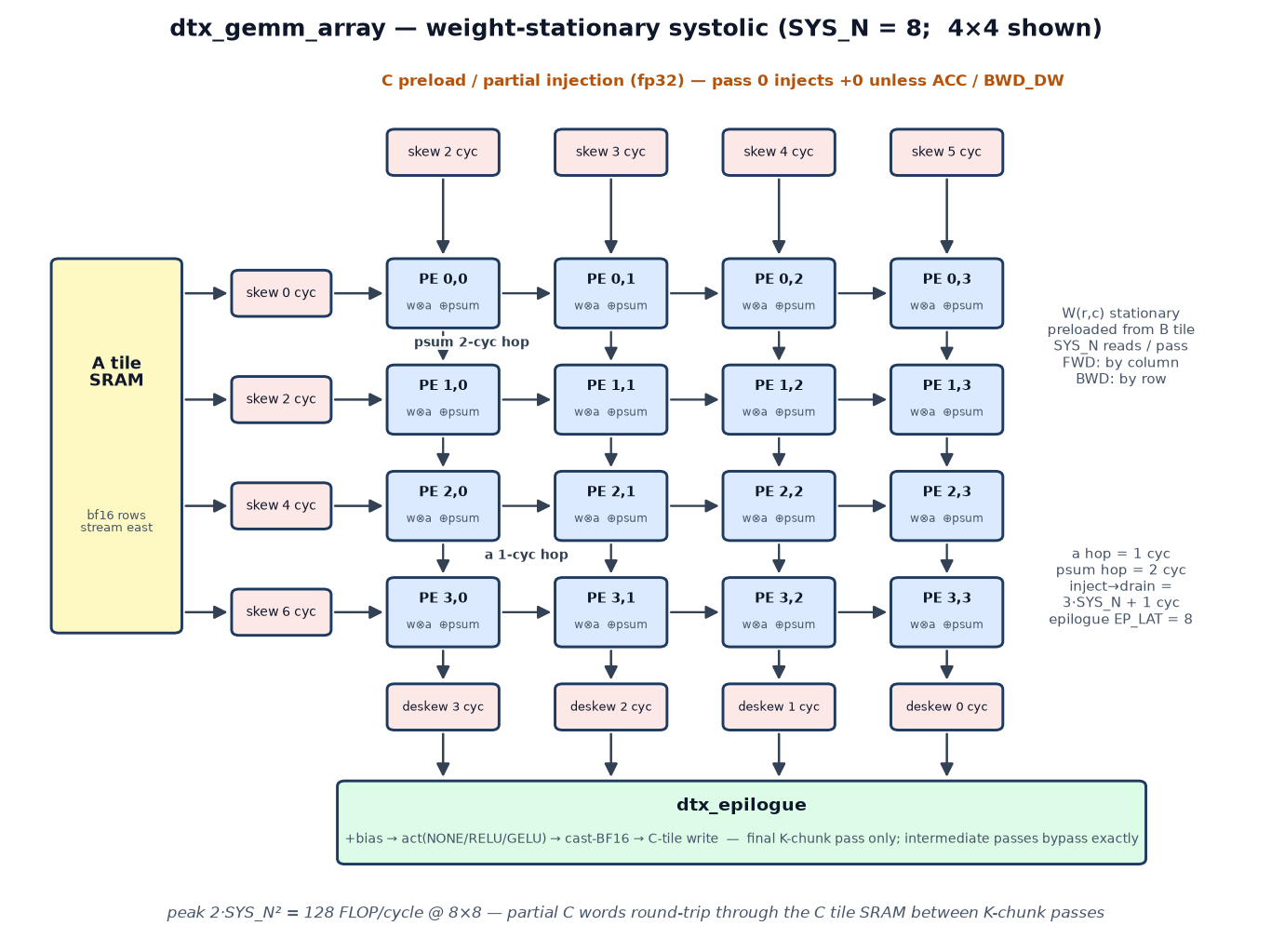}
  \caption{\code{dtx\_gemm\_array}: weight-stationary PEs; $A$ enters from
  the west skewed $2r$ cycles for row $r$, C partials/preloads enter from
  the north skewed $c{+}2$ cycles for column $c$, and the south edge
  deskews by $\code{SYS\_N}{-}1{-}c$, so a full C row word exits
  $\code{ARR\_LAT}=3\cdot\code{SYS\_N}{+}1$ cycles after its $A$ word is
  injected.}
  \label{fig:systolic}
\end{figure}

\code{dtx\_gemm\_array} is a $\code{SYS\_N}\times\code{SYS\_N}$
weight-stationary array~\cite{jouppi2017tpu} computing the generic form
$C[M,N] \mathrel{(+)=} A[M,K]\cdot B[K,N]$ (Fig.~\ref{fig:systolic}).
PE$(r,c)$ holds a stationary BF16 weight; each PE is exactly one
\code{dtx\_bf16\_mul} (2-stage, exact) plus one southbound
\code{dtx\_fp32\_add} (2-stage), i.e.\ 2\,FLOP/PE/cycle. The timing closes
by a self-alignment identity rather than by handshakes: the eastbound $A$
hop is 1~cycle and the southbound partial-sum hop is 2~cycles (the adder's
own pipe), so with the west edge injecting row $r$ delayed by $2r$ cycles
(matching the 2-stage multiplier), PE$(r,c)$'s product arrives at its adder
\emph{exactly} when PE$(r{-}1,c)$'s partial sum does. There is no
accumulate register beyond the adder's output register, and no valid
crossbar inside the array, one shared skew line carries validity.

The north edge closes the same identity for row~0: the C word injected into
column $c$ is delayed $c{+}2$ cycles to meet row~0's product. The south
edge deskews column $c$ by $\code{SYS\_N}{-}1{-}c$ registers, so drained C
words are row-aligned; total injection-to-drain latency is
$\code{ARR\_LAT}=3\cdot\code{SYS\_N}+1$ (25 at $\code{SYS\_N}{=}8$, 13 at
4). A meta pipe of depth $\code{ARR\_LAT}{+}1$ carries each beat's C write
address/bank from issue to drain, and a simulation assertion pins
\code{dr\_valid} to it every cycle.

\emph{Arbitrary dims without an ordering contract.} Dimensions $1..64$ run
as $\lceil N/\code{SYS\_N}\rceil \times \lceil K/\code{SYS\_N}\rceil$
passes; partial sums \emph{round-trip through the C tile SRAM}: each pass
re-injects the previous partial C word at the north edge and drains the
updated word at the south edge, so the engine sustains II${=}1$ (one $A$
row word per cycle in \code{S\_STREAM}) at any $K$. Pass~0 injects $+0$
unless the op accumulates (\code{G\_BWD\_DW}, or \code{FLG\_G\_ACC}), in
which case the preloaded C is injected. The resulting accumulation
order, down the column within a $K$-chunk, chunk by chunk across
passes, is a \emph{fixed pipeline order with no sequential-sum contract};
Sec.~\ref{sec:numerics} makes this the verification premise. The per-pass
FSM is \code{LOAD\_W} ($\code{SYS\_N}$ weight-word reads plus an optional
bias word, $\code{SYS\_N}{+}2$ cycles) $\to$ \code{S\_STREAM} ($M$ beats,
II${=}1$) $\to$ \code{S\_DRAIN} (counts epilogue write completions via
\code{ep\_wr\_done}); an op costs
$\approx N_C K_C (M + 4\,\code{SYS\_N} + 11)$ cycles.

\emph{Dataflows and ragged edges.} \code{G\_FWD} ($Y{=}XW^{\mathsf T}$)
column-loads weights (B-tile row $n_0{+}c$ becomes array column $c$);
\code{G\_BWD\_DX} ($dX{=}dY\,W$) row-loads them; \code{G\_BWD\_DW}
($dW \mathrel{+}= dY^{\mathsf T}X$) row-loads with C preload forced on.
(The $dY^{\mathsf T}$ operand is staged by the host in v1, the DMA's 2-D
stride is a row-major mirror and cannot synthesize a transpose;
Sec.~\ref{sec:verif}.) Ragged lanes cost nothing numerically:
out-of-range weights load as $+0$ and out-of-range $A$ lanes inject $+0$,
and since $\code{fp32\_add}(x,+0)=x$ exactly, short-$K$ results are
bit-identical to full-width ones with zero padding; out-of-range C lanes
are written with don't-care data that the DMA never stores.

\subsection{Fused drain epilogue}

\code{dtx\_epilogue} sits between the array's drain stream and the C-tile
write port: $\code{EP\_LAT}=8$ cycles ($=$ bias add 2 $+$ activation 5 $+$
cast 1), $\code{SYS\_N}$ FP32 lanes, II${=}1$, free-running, bias,
activation, and BF16 cast cost \emph{no extra memory pass}. Stage~A is one
\code{dtx\_fp32\_add} per lane against the bias word (BF16 lanes widened
during \code{LOAD\_W}), with a 2-register \emph{exact raw bypass} in
parallel. Stage~B evaluates \code{dtx\_gelu\_pwl} (5-stage) with the RELU
path (the package sign-mux, not an FP op) and the identity path
delay-matched in 5-deep registers, so all three activations have identical
latency. Stage~C optionally narrows: the result is stored as
\code{\{cast\_bf16(x), 16'h0\}} so the C tile remains uniform FP32 lanes
and the DMA takes bits $[31{:}16]$ on a BF16 store. The load-bearing bit is
\code{in\_final}, which rides per word: intermediate $K$-chunk passes of
the SRAM round-trip traverse the epilogue through the exact bypasses
(bit-preserving), and bias/activation/cast apply only on the last $K$ chunk
of an output chunk. Config inputs are quasi-static, legal to change only
while the pipe is empty, which the fully-drained pass structure guarantees.

\subsection{Vector processing unit}

\code{dtx\_vpu} is a \code{LANES}-wide FP32 engine with twelve opcodes
(\code{V\_ADD}, \code{V\_MUL}, \code{V\_CAST}, \code{V\_ACT\_BWD},
\code{V\_MSE\_GRAD}, \code{V\_CE\_GRAD}, \code{V\_NOISE},
\code{V\_BIAS\_BWD}, \code{V\_LN\_FWD/BWD}, \code{V\_SOFTMAX\_FWD/BWD}),
each II${=}1$ per \code{LANES}-element group. Rather than per-op datapaths,
every pass is a static configuration of one \emph{canonical lane pipe}
\[
[\mathrm{gelu}']{\to}\{\mathrm{M1}{\parallel}\mathrm{M2}\}{\to}
\mathrm{A1}{\to}\mathrm{A2}{\to}\mathrm{M3}{\to}[\mathrm{EXP2}]
{\to}\mathrm{M4}{\to}\mathrm{A3},
\]
whose operand muxes select among the three read-port operands (with static
delay taps), the scalar registers $S_0..S_2$, and exact identities
($\times 1.0$, $+0$); reduction taps sit at $\{v_a,\mathrm{M1},\mathrm{M3},
\mathrm{EXP}\}$ and write taps at any pipe station. Multi-pass ops sequence
per row: LN forward runs mean, variance$\to$\code{rstd}, normalize
($\pm\gamma/\beta$ affine), and an optional $\hat x$/stats cache pass for
the backward; softmax forward runs max, $\sum \mathrm{exp2}((x{-}m)\log_2
e)$ with a reciprocal, then scale. Between passes, shared scalar units (one
mul, one add, \code{dtx\_recip\_nr}, \code{dtx\_rsqrt\_nr}) fold the row
reduction into $S_0..S_2$; $1/N$ itself is computed once per op from an
exact integer-to-FP32 conversion of \code{cols}.

Reductions are two-level trees: a \code{LANES}-input \code{dtx\_tree\_add}
/ \code{dtx\_tree\_max} consumes one group per cycle; group results collect
into a per-row buffer, and a second $\code{VPU\_GMAX}$-input tree
(\code{TILE\_N}/\code{LANES} $=8$ inputs) fires once per row, a
\emph{re-tree}, never a loop-carried chain. Tail lanes
($\code{cols}\bmod\code{LANES}$) are identity-padded ($+0$ into the sum
tree, $-\infty$ into the max tree) and byte-enable-masked at the write
port, so partial groups are exact.

\code{V\_BIAS\_BWD} (column sums across rows) is the one op that wants an
accumulator, and it gets one without a loop-carried add: a
\emph{2-bank, row-parity} FP32 accumulator per (group, lane). Bank $=$ row
LSB, so a given (bank, group) cell recurs every $2G \ge 2$ issue
cycles, exactly the adder latency, and the $G{=}1$ back-to-back case
forwards the adder output combinationally. A final pass emits
$db = \mathrm{ACC}_0 + \mathrm{ACC}_1$ through the same tap network.

\subsection{Fused AdamW optimizer}

\code{dtx\_opt} instantiates \code{LANES} identical, fully-pipelined AdamW
lanes: $m, v, w, g$ stream in and $m', v', w', \code{bf16}(w')$ stream out
at II${=}1$ element/lane/cycle. The pipeline map (cycles after
\code{in\_valid}) is: $t{+}2$ five parallel multiplies ($\beta_1 m$,
$(1{-}\beta_1)g$, $g^2$, $\beta_2 v$, $\mathrm{lr\_wd}\cdot w$); $t{+}4$
$m'$; $t{+}6$ $v'$ and $\hat m = m' r_{\beta_1 k}$, the early output group
$\code{LAT\_OPT\_MV}{=}6$; $t{+}8$ $\hat v$ and $\mathrm{lr}\cdot\hat m$;
$t{+}33$ \code{rsqrt\_nr}$(\hat v)$; $t{+}35$ the multiply
$\hat v\cdot\mathrm{rsqrt}(\hat v)$ forming $\sqrt{\hat v}$,
with a delay-matched special bypass replicating
\code{sqrt\_f32} ($\hat v{=}0 \to 0$, not $0\cdot\infty{=}$NaN, zero-init
state is the common case); $t{+}37$ $\mathrm{den} = \sqrt{\hat v} +
\varepsilon$ (PyTorch placement, outside the root); $t{+}56$
\code{recip\_nr}; $t{+}58/60/62$ update, subtract, decoupled decay;
$t{+}63$ output register with the single BF16 narrowing, 
$\code{LAT\_OPT\_W}=63$, dominated by the Newton units ($25{+}19$).
Operands that must meet late results ride plain delay lines ($w$ 58 deep,
$\mathrm{lr}\cdot\hat m$ 48 deep); decay on the \emph{pre-update} weight
uses the $t{+}2$ product delayed 58 cycles, and \code{adamw\_en}${=}0$
subtracts $-0.0$ through the same adder so both settings are
latency-identical and exact. $(1{-}\beta_{1,2})$ are formed on chip by two
free-running pipelined adders; all hyperparameters are full 32-bit
CSR-backed FP32 immediates, never the 16-bit slot immediate, which
forecloses a silent-truncation hazard (a hyperparameter narrowed in flight,
and mirrored identically by a reference, passes every per-op check while
training wrong; Sec.~\ref{sec:arch:csr}). A sticky
\code{nan\_seen} flags any non-finite $g$ lane.

The slot-2 wrapper \code{dtx\_opt\_eng} streams the four tiles from the
unified tile space and updates \emph{in place}: writes trail reads by the
fixed pipe latency, so no copy is needed. Its placement contract ($m, v, w$
in three different regions; $g$ sharing with one) is exactly the crossbar
budget below. Abort gates all writes, flips the RNG epoch, and drains for a
bounded $80 > \code{LAT\_OPT\_W}$ cycles.

\subsection{Pipelined Philox and the banked ICDF}

\code{dtx\_philox\_pipe} is a fully pipelined
Philox-4$\times$32-10~\cite{salmon2011philox}: one
canonical round per stage ($R{=}10$), throughput one 128-bit block per
cycle, the per-stage key schedule $k_r = \mathrm{key} + rW$ derived
combinationally from the quasi-static seed, a shadow assertion pinning
every emitted block to \code{philox\_block()}, and a 1-bit epoch tag whose
mismatch discards in-flight blocks after an abort. The pipelining is the
design law of Sec.~\ref{sec:numerics} at work: a combinational equivalent
(80 $32{\times}32$ multiplies per beat) is un-mappable at chip scale.

\code{dtx\_rng} wraps $\lceil\code{LANES}/4\rceil$ pipes; lane $l$ consumes
word $l\bmod 4$ of pipe $\lfloor l/4\rfloor$, and the wrapper advances the
block counter by \code{NPIPES} per beat, so lane $l$ of beat $b$ is the
unique global Philox word $4(\mathrm{ctr}_0 + b\,\code{NPIPES} + \lfloor
l/4\rfloor) + (l\bmod 4)$, a pure function of (seed, stream, counter),
reproducible across stalls and replays. Gaussians use a 512-entry FP32
probit ROM (entry $i = \Phi^{-1}((i{+}0.5)/512)$, Acklam-generated) with
linear interpolation: index $u[31{:}23]$, fraction $u[22{:}13]$, three
pipelined FP ops (subtract, multiply, add); index 511 clamps; the
antithetic option flips the sign on $u[0]$, a bit the index/fraction never
consume. The ROM is replicated per two lanes so each hardened instance
serves four reads per cycle. Uniforms take a matched-latency bypass
computing $u[31{:}8]\cdot 2^{-24}$ \emph{exactly} (24 bits fit the FP32
significand), which is why the scoreboard holds uniforms to bit-equality.
End to end, $\code{LAT\_RNG} = 10 + 8 = 18$ and the engine emits
\code{LANES} values per cycle. The quantized generator's analytic law
(exact integration over the index/fraction grid) has mean $+5.9{\times}
10^{-3}$ (the clamp's bias), variance $0.9973$, support $|z|\le 3.0973$.

\subsection{DMA and burst engines}

The slot-3 DMA front-end walks one 2-D strided op at a time
(\code{D\_LD\_TILE}, \code{D\_ST\_TILE}, \code{D\_LDTDR}), converting rows
into $\{$address, bytes$\}$ commands. Memory rows live at $\mathrm{base} +
r\cdot\mathrm{pitch}$ with arbitrary byte alignment; tile rows are
word-aligned, packed $\lceil \mathrm{cols}\cdot\mathrm{esize}/32\rceil$
words per row with the tail word byte-enabled, at a tile-word pitch carried
in \code{imm[15:11]} (the GEMM-consumable layouts are pitch 4 for BF16 rows
and 8 for FP32 rows). \code{dtx\_dma\_rd} splits commands into INCR bursts
($\le$16 beats, 4\,KB-safe), keeps four in flight on a single ID, and packs
arbitrarily-aligned read data into a contiguous byte stream at one 256-bit
beat per cycle; its error is a self-clearing one-shot pulse.
\code{dtx\_dma\_wr} is upgraded for throughput: it does \emph{not} wait for
\code{BRESP} between bursts, a 4-credit counter gates \code{AWVALID}
while write data stays strictly serialized behind its own \code{AW}, so
AXI4 write ordering holds trivially; \code{WSTRB} masks partial beats, and
abort completes every issued burst with \code{WSTRB}${=}0$ padding before
going idle. Its error, by contrast, is a \emph{sticky level} cleared at op
start, the asymmetry behind the one RTL bug the program found (B2, a
recovery race re-latching the prior op's error;
Sec.~\ref{sec:verif}). The tile-read path (issue / SRAM latency / 2-deep
skid) sustains one word per cycle whenever AXI keeps up, and the perf
counters that prove compute/DMA overlap (Sec.~\ref{sec:results}) count
these beats.

\subsection{Unified tile space and crossbar}

All engines share one 64\,KB byte-addressed tile space, physically three
\code{dtx\_tbmem} regions: A (16\,KB @ \code{0x0000}), B (16\,KB @
\code{0x4000}), C (32\,KB @ \code{0x8000}), each two ping-pong banks of
256-bit words with two independent 1-cycle synchronous read ports and one
byte-enabled write port (read-first, \code{\$mem}-inferable, array
unreset). A fixed-priority crossbar maps 11 read clients (GEMM
A/B/C, hard-wired to their regions, DMA, VPU A/B/C, OPT $m/v/g/w$) onto
2 read grants per region per cycle, and 6 write clients (epilogue, DMA,
VPU, OPT $m'/v'/w'$-or-RNG) onto 1 write grant per region per cycle.
Return routing is a pair of \emph{registered selects} per client
(region, port captured on grant), which works because every DTX engine
consumes read data on the landing cycle $T{+}1$, port sharing can
therefore never corrupt a granted read. Over-subscription is not
arbitrated: a dropped request is a \emph{program error}, \code{\$error}-ed
in simulation and deterministic garbage in hardware; the VLIW
\code{wait\_mask} (Sec.~\ref{sec:arch}) exists precisely so the compiler
schedules slot concurrency inside the $2R{+}1W$ budget, e.g.\ the AdamW
placement contract above, or LN-affine's rule that $\gamma/\beta$ must not
share a region with the input tile.

\section{Verification}\label{sec:verif}

The numerics of Sec.~\ref{sec:numerics} dictate the methodology. DTX
contracts no reduction order, its reductions are pipelined adder trees in the
VPU and interleaved pipelined accumulators in the systolic K-loop, so the
only promise the ISA makes about a reduction is its \emph{accuracy}, not its
bit pattern. The verification environment is therefore built around a different
oracle: a full-UVM testbench (\code{verif/testbench/}, one class per file,
per-agent directories, base/extend/stress sequence libraries) whose golden
model computes every operation in FP64 (SystemVerilog \code{real}) and whose
scoreboard checks every element the program wrote, with \code{===} equality
exactly where the architecture leaves no arithmetic latitude, and with a
derived error bound everywhere else.

\subsection{Why an FP64 oracle with tolerance is the right contract}
\label{sec:verif:oracle}

A bit-exact reference for DTX would have to reproduce the netlist's exact
reduction topology: the tree pairing order of \code{dtx\_tree\_add}, the
interleaving of the systolic column accumulators, the epilogue's register
bypass for intermediate $K$-chunk passes. Such a reference verifies the
netlist against a transcription of itself, it can catch copying errors but
not wrong mathematics, and every micro-architectural change (a different
$\lceil K/8\rceil$ chunking, a deeper skid buffer) would break it without any
functional regression. Conversely, checking against a straightforward
sequential-sum software model would fail spuriously: for a depth-$d$
($d=\log_2 K$) tree the classical rounding bound is
$|\mathrm{fl}(\Sigma)-\Sigma| \le d\,u\sum_i |x_i|$ with $u=2^{-24}$ the FP32
unit roundoff, versus $(K{-}1)\,u\sum_i|x_i|$ for the sequential chain, the
tree is generally \emph{more} accurate, but bit-different. The correct oracle
for unordered numerics is a reference of much higher precision plus an error
budget derived from the rounding model. The golden model computes in FP64
($u_{64}=2^{-53}$, $2^{29}\times$ finer than the FP32 target), and the
scoreboard (\code{dtx\_scoreboard.sv}) applies per element
\begin{equation}\label{eq:tol}
|x_\mathrm{dut}-x_\mathrm{ref}| \;\le\;
  s\!\left[\left(10^{-6}+a_x\right)
  + \left(r_\mathrm{base}+r_x\right)|x_\mathrm{ref}|\right],
\end{equation}
\begin{equation}\label{eq:rtol}
r_\mathrm{base} = 2^{-p}\!\left(1+\tfrac{\log_2 K}{8}\right),\qquad
p=\begin{cases}7 & \text{BF16-stored,}\\[1pt]
              20 & \text{FP32,}\end{cases}
\end{equation}
where $K$ is the per-tensor reduction depth attached by the model to each
written element, $a_x$/$r_x$ are per-tensor extras the model computes, and
$s$ is an environment knob (default 1.0). Every constant is accounted for:

\begin{itemize}
\item \textbf{FP32 base $2^{-20}=16u$} is a depth-independent floor that
  absorbs a handful of chained FP32 operations (epilogue bias add, cast
  chains) plus mild $\sum|x_i|/|\Sigma|$ amplification.
\item \textbf{The depth term} $(1+\log_2 K/8)$ adds $2^{-20}/8 = 2u$ of
  relative budget per tree level, a linear-in-depth allowance with a
  2-ULP-per-level slope, i.e.\ $2\times$ margin over the 1-ULP-per-level
  worst case of the tree bound above. At the maximum tile depth $K{=}64$ the
  factor is 1.75, giving $r_\mathrm{base}\approx 1.67\times10^{-6}$.
\item \textbf{BF16 base $2^{-7}$}: BF16 (1-8-7) unit roundoff is $2^{-8}$;
  a BF16-stored result's error is dominated by the single final
  round-to-nearest-even narrowing ($\le 2^{-8}$ relative), and $2^{-7}$ is
  twice that, covering the FP32$\to$BF16 double rounding of an
  already-rounded FP32 value. The same depth shape keeps the underlying FP32
  tree covered uniformly.
\item \textbf{$a_\mathrm{tol}=10^{-6}$} is the absolute floor where
  $|x_\mathrm{ref}|\approx 0$: the DUT flushes subnormals to zero
  (Sec.~\ref{sec:numerics}) and near-total cancellation makes relative error
  meaningless there.
\item \textbf{Per-tensor extras} $a_x$/$r_x$ carry the \emph{documented} PWL
  approximation error of the transcendental units into tensors that used
  them (\code{exp2} $\le 1.2\times10^{-5}$ rel., GELU
  $\le 2.3\times10^{-4}$ abs., GELU$'$ $\le 8\times10^{-4}$ abs.) plus
  cancellation slack the model detects while tracking $\sum|\cdot|$ per
  output element.
\end{itemize}

\emph{The \code{EXACT} subset.} Tolerance is applied only where arithmetic
reordering exists. Where the architecture defines the bits, the scoreboard
compares bits: DMA loads/stores (pure byte copies), casts of exact values,
Philox output (an integer datapath, and the uniform conversion
$u[31{:}8]\cdot 2^{-24}$ is exact), the Gaussian draws (the reference mirrors
the ICDF LUT and the identical three-operation FP32 interpolation chain), and
all CSR/TDR/IRAM readback. In the 17-test regression the exact class is not
a corner: the MLP training acceptance test alone checks 28{,}816 exact
elements alongside 78{,}292 tolerance elements.

\emph{The bound has teeth.} Bug B3 (Sec.~\ref{sec:verif:bugs}) demonstrated a
measured $5{,}340\times$ violation of Eq.~\eqref{eq:tol} the moment a program
broke the bound's premise, while all 78{,}292 tolerance-checked elements of
the compliant training suite pass with zero failures. The budget is tight
enough to catch a single mis-rounded input, not a vacuous ``anything
goes'' band.

\subsection{Environment architecture}\label{sec:verif:env}

The scoreboard is fully monitor-driven: it reconstructs device state from
the AXI-Lite transaction stream, i.e.\ from exactly what host software can
observe. Every monitored CSR write updates the golden model's CSR
mirror, IRAM and TDR windows byte-strobe-accurate, PC, seeds,
\code{RNG\_CTR}, and the eight full-32-bit AdamW hyperparameter registers
(the discipline of Sec.~\ref{sec:arch:csr}: no hyperparameter ever rides a
truncated immediate). When the monitor observes \code{CTRL.START}, the golden model
(\code{dtx\_golden\_model.sv}, 1{,}057 lines) executes the entire VLIW
program in zero simulation time against two \emph{mirror spaces}
(\code{dtx\_ref\_space.sv}): a 64\,KB image of the unified tile space, and a
private snapshot copy of the DUT-visible external memory. Execution
publishes a per-slot decode trace to the coverage model
(Sec.~\ref{sec:verif:covmodel}) and a program-result prediction (clean halt,
or fault with \{\code{pc},\,\code{slot},\,\code{err}\}).

\emph{Record classes.} Each element the model writes is recorded at its
byte address as either \code{REF\_EXACT} (the bit image is authoritative;
compared with \code{===}) or \code{REF\_REAL} (an FP64 value plus the
metadata of Eq.~\eqref{eq:tol}: reduction depth $K$, BF16-precision flag,
per-tensor extras). Addresses with \emph{no} record are never
checked, uninitialized tile SRAM and the architecturally don't-care
ragged-edge lanes of a partial GEMM tile stay out of the oracle rather than
being laundered into it.

\emph{Whole-space sweep.} The status poll that observes \code{DONE|ERR}
(the same poll discipline the driver uses) triggers the end-of-program
check: the predicted outcome is compared first, and on a predicted-clean
program the scoreboard sweeps \emph{every recorded element of both spaces},
reading DUT tile bits through a read-only hierarchical backdoor
(\code{dtx\_probe\_if}, the environment's one white-box concession) and
external bytes from the memory model. Any \code{X} in swept DUT data is a
failure; NaN expectations must be matched by NaN; infinities must match in
sign. A stray write anywhere in 64\,KB of tile space or in the touched
external ranges, or a missing one, fails the test. After a passing sweep
the model adopts the DUT's stored bits as the new reference image for
tolerance-class external tensors (\code{sync\_ext\_real}), so multi-program
tests re-baseline instead of compounding tolerance across programs.

\emph{Validation mirror.} The golden model reimplements the sequencer's
\code{S\_CHECK} validation stack, opcode legality, flag encodings,
per-operand TDR checks, and the 48-bit footprint columns, with identical
error codes and the same fault-slot selection, followed by the per-engine
defensive decode checks as runtime faults. Fault programs are therefore
\emph{predicted}, not tolerated: the scoreboard demands \code{STATUS.ERR}
with the right \code{CAUSE}\,\{\code{pc},\,\code{slot},\,\code{err}\}, and a
model-predicted fault that the DUT completes ``cleanly'' is equally an
error. AXI-fault programs are predicted too: the model imports the
responder's error-injection ranges at \code{START}. On \code{CTRL.ABORT}
the model deliberately drops all data expectations (partial writes are
timing-dependent), predicts \code{ERR\_ABORT}, and invalidates its
\code{RNG\_CTR}/\code{nan\_seen} predictions until software rewrites
them, abort recovery is then proven by the fully checked program that
follows.

\emph{Agents.} Three agents drive and observe the DUT's complete pin-level
contract. The \code{axil\_agent} masters the CSR port. The
\code{axi\_mem\_agent} is a reactive 256-bit-data/40-bit-address AXI4 slave
BFM over a byte-sparse memory model: independent read and write channels,
INCR bursts to 16 beats, randomized handshake latencies and ready duty
cycles, kill-and-restart reset behavior, and \code{SLVERR} injection over a
configured address range, the mechanism behind the DMA fault tests and the
discovery of B2. The \code{irq\_agent} observes the interrupt pin itself
(an audit rule of the environment: every output pin must be seen by a
checker), and the
scoreboard also enforces a perf-counter sanity invariant: every observed
per-slot \code{BUSY} value must be $\le$ any \code{PERF\_CYCLES} value read
no earlier, both being monotonic while a program runs.

\subsection{Functional coverage model}\label{sec:verif:covmodel}

Functional coverage (\code{dtx\_env\_cov.sv}) samples the golden model's
\emph{replayed} slot trace, the decode of what the IRAM actually
contained, rather than what a sequence intended to write. This
stimulus-side sampling guards against a coverage-vacuity hazard: a DUT and
reference can happily agree on a corrupted program, and only coverage of the
replayed descriptor fields exposes it. Seven covergroups cover the
ISA and the outcome space: \code{cg\_gemm} (dataflow $\times$ activation
$\times$ bias/accumulate/cast epilogue-flag crosses, dimension corner bins),
\code{cg\_vpu} (all 12 ops $\times$ destination format, LN affine/cache
variants, cast direction, tail-lane column bins), \code{cg\_opt} (AdamW
decay flag, RNG gauss/uniform $\times$ format $\times$ antithetic),
\code{cg\_dma} (op $\times$ format $\times$ row-pitch encodings),
\code{cg\_word} (all 15 co-issue slot combinations, loop open/close/single
and count bins, halt), \code{cg\_wait} (\code{wait\_mask} patterns crossed
with the consuming slot), and \code{cg\_result} (done vs.\ error, all six
error codes crossed with the faulting slot).

\subsection{The 17-test suite}\label{sec:verif:tests}

Table~\ref{tab:tests} lists the regression suite
(\code{verif/test\_plan.csv}). The 2026-07-07 regression is 17/17 green:
\code{UVM\_ERROR}${}={}$\code{UVM\_FATAL}${}={}$\code{UVM\_WARNING}${}=0$,
every test runs to \code{\$finish}, and the scoreboard reports zero failing
elements across all sweeps.

\begin{table}[t]
  \centering
  \caption{The 17-test regression suite. Test names are
  \code{dtx\_\emph{name}\_test}; all pass with zero scoreboard failures.}
  \label{tab:tests}
  \footnotesize
  \begin{tabularx}{\columnwidth}{@{}l l X@{}}
    \toprule
    Test & Type & Purpose / key checks \\
    \midrule
    \code{base} & bring-up & env/agent/vif hookup; CSR ID; STATUS clean out
      of reset \\
    \code{gm\_sanity} & bring-up & golden-model wiring: one program
      touching 3 engines and all 3 tile regions; \code{RNG\_CTR} advance \\
    \code{smoke} & basic & first full path: DMA
      $\to$ GEMM (\code{relu}, bias) $\to$ DMA store; done-IRQ on the pin \\
    \code{gemm\_dataflow} & feature & FWD / BWD-$dX$ / BWD-$dW$ on ragged
      odd shapes ($13{\times}11{\times}20$, $16{\times}15{\times}17$,
      $10{\times}12{\times}9$ with FP32 accumulate) \\
    \code{gemm\_epilogue} & feature & fused-drain cross
      \{bias\}$\times$\{none, relu, gelu\}$\times$\{cast\}: 14 programs
      incl.\ negative pre-activations \\
    \code{gemm\_multitile} & feature & $64^3$ max tile; loop-driven
      double-buffered stream; perf-counter overlap proof
      (Sec.~\ref{sec:results:overlap}) \\
    \code{vliw\_loop} & control & zero-overhead loops: counts 1/2/255,
      multi-slot body, HALT-in-loop priority, \code{BAD\_PC} fall-off
      fault \\
    \code{vliw\_coissue} & control & serial vs.\ all-4-slot word vs.\ RAW
      chain; $\mathrm{cyc}_\mathrm{co} < \mathrm{cyc}_\mathrm{serial}$; all
      per-slot busy counters nonzero \\
    \code{opt\_adamw} & feature & fused AdamW: $k{=}1$ decay-on, $k{=}100$
      decay-off, NaN-gradient lane; \code{nan\_seen} set/W1C (found B1) \\
    \code{rng} & feature & gauss/uniform $\times$ fp32/bf16 $\times$
      antithetic, all \code{EXACT}; 98{,}304-sample moment check; counter
      continuation across re-issues \\
    \code{dma} & feature & odd-pitch 2-D loads/stores, \code{LDTDR} +
      readback; \code{SLVERR} read and write faults each followed by a clean
      re-run (found B2) \\
    \code{vpu\_map} & feature & 9 map ops with tail lanes
      (ADD/MUL/CAST/ACT-BWD/MSE-GRAD/NOISE), both write formats \\
    \code{vpu\_norm} & feature & LayerNorm fwd affine/cache variants;
      fwd$\to$bwd program consuming the cached $\hat{x}$/rstd \\
    \code{vpu\_softmax} & feature & softmax fwd ($\pm 25$ logits) and bwd,
      CE-grad, column tree-sum bias-bwd \\
    \code{err} & error & every error code from Sec.~\ref{sec:arch}, aborts
      mid-GEMM and mid-VPU, then a fully checked recovery program \\
    \code{train\_attn} & acceptance & one attention block
      (LN $\to QK^{\mathsf T}\!\to$ softmax $\to\cdot V\to$ proj $\to$
      residual) fwd + bwd, host-orchestrated stages \\
    \code{train\_mlp} & acceptance & 20-iteration diffusion-MLP training;
      \emph{semantic gate: the loss must decrease}
      ($56.4\!\to\!26.0$); 107{,}108 elements checked \\
    \bottomrule
  \end{tabularx}
\end{table}

\emph{The two semantic acceptance gates.} The last two rows institutionalize
a principle no equivalence check can supply: an equivalence check
cannot catch a defect its reference shares, so \emph{the workload is the
final oracle}, and it belongs in the regression, not in a post-silicon
bring-up plan. \code{dtx\_train\_mlp\_test} runs a complete
diffusion-denoiser training step twenty times on the device: Gaussian noise
draw, closed-form noising, forward GEMM with fused bias/ReLU, MSE gradient,
activation backward, both backward GEMMs, bias gradient, and fused AdamW on
weights and bias, with the RNG counter reset per step so the noise, and
hence the loss trajectory, is deterministic. The test asserts
$\overline{\mathcal{L}}_{0..2} > \overline{\mathcal{L}}_{17..19}$, a
\emph{training-loss-decreases} regression check; the measured trajectory
falls from 56.44 to 26.02 (Sec.~\ref{sec:results:train}). Every one of the
100 constituent programs is simultaneously swept against the FP64 oracle
(28{,}816 exact + 78{,}292 tolerance elements, zero failures), so the gate
is semantic \emph{and} numeric at once. \code{dtx\_train\_attn\_test} does
the same for a transformer self-attention block, forward and backward,
exercising LayerNorm, all three GEMM dataflows, softmax forward/backward,
and the residual path on real attention data. Both tests are
host-orchestrated as stage programs, a structure forced by finding B3, as
described next.

\subsection{Coverage results and closure plan}\label{sec:verif:cov}

\begin{figure}[t]
  \centering
  \includegraphics[width=\columnwidth]{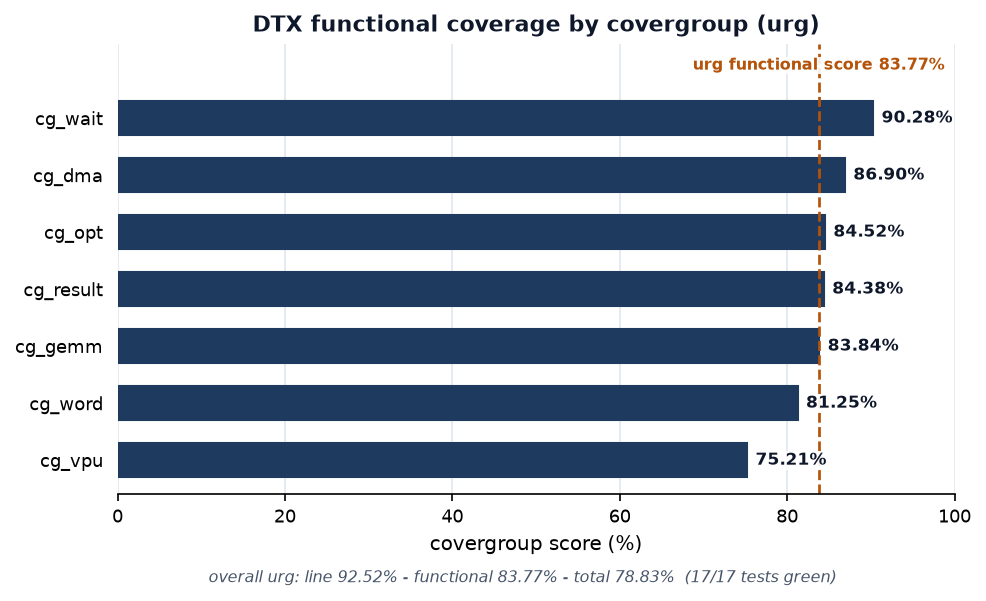}
  \caption{Per-covergroup functional coverage (\code{urg}). The lowest
  group, \code{cg\_vpu} (75.21\%), is dominated by the op $\times$ column
  cross; the group score is 83.77\%.}
  \label{fig:cov}
\end{figure}

Table~\ref{tab:cov} and Fig.~\ref{fig:cov} report the first coverage pass
(\code{urg}, 18 merged databases). This is deliberately reported as a
\emph{first pass}: no exclusion (waiver) files have been authored yet, and
we make no closure claim.

\begin{table}[t]
  \centering
  \caption{First-pass coverage (\code{urg}, 18 merged databases,
  17/17 tests green). No exclusions applied; closure is planned, not
  claimed.}
  \label{tab:cov}
  \footnotesize
  \begin{tabularx}{\columnwidth}{@{}l r X@{}}
    \toprule
    Metric & Score (\%) & Dominant residue \\
    \midrule
    Line      & 92.52 & \code{dtx\_vpu} multi-pass FSM arms \\
    Condition & 67.01 & wide guard expressions, partial minterms \\
    Branch    & 80.79 & error/abort arms of engine FSMs \\
    Toggle    & 79.33 & unused CSR register bits (\code{dtx\_csr}
                        toggle 28\%) \\
    FSM       & 49.66 & \code{dtx\_vpu} (39/73 transitions uncovered);
                        \code{dtx\_dma} error$\to$error edges
                        (mostly unreachable) \\
    Assert    & 98.73 &, \\
    Functional (groups) & 83.77 & crosses, Fig.~\ref{fig:cov} \\
    \midrule
    \textbf{Total score} & \textbf{78.83} & \\
    \bottomrule
  \end{tabularx}
\end{table}

The holes decompose into three kinds. (i)~\emph{Reachable-by-plan}: the
four largest functional crosses, error code $\times$ faulting slot
(37.5\%), GEMM dataflow $\times$ epilogue flags (41.7\%), VPU op $\times$
column bins (47.9\%), and co-issue slot combinations (50\%), plus the
\code{dtx\_vpu} FSM transitions. Four closure tests are already specified
in \code{test\_plan.csv} (\code{dtx\_cov\_vpu\_fsm\_test},
\code{dtx\_cov\_err\_slot\_test}, \code{dtx\_cov\_gemm\_flags\_test},
\code{dtx\_cov\_coissue\_mask\_test}), each a directed sweep of the named
bins with the standard golden sweep as checker. (ii)~\emph{Waiver
candidates}: the \code{dtx\_dma} error-latch FSM's error$\to$error edges are
mostly unreachable by construction, and \code{dtx\_csr} toggle residue is
unused register bits; these require reviewed exclusion files (six are
planned) before any strict-mode figure is quoted. (iii)~\emph{Stimulus
gaps}: \code{dtx\_dma\_wr}'s drain state needs a mid-store abort, which the
planned error-slot test provides. Until (i)--(iii) land, the honest summary
is the raw table above.

\subsection{Bugs found: B1--B3}\label{sec:verif:bugs}

Five parallel test-development tracks (GEMM, VPU, OPT/RNG/DMA, VLIW
control, training acceptance) yielded three catalogued defects
(\code{docs/learnings.txt}); the reconciled full regression re-confirms all
three fixed and found no fourth RTL or golden-model defect (the later
PD-configuration bring-up caught one verification-\emph{stimulus} bug,
Sec.~\ref{sec:results:synth}). One is an RTL bug, one a golden-model bug,
one an architectural-methodology finding, a distribution that itself says
something about where tolerance-based verification bites.

\emph{B1 (golden model): \code{\$sqrt(NaN)} aborts the simulator.} The
spec-required non-finite-gradient scenario (which must set
\code{OPT\_STAT.nan\_seen}) drove the reference AdamW's bias-corrected
second moment $\hat v$ to NaN; the guard tested only $\hat v<0$, which is
\emph{false} for NaN, so control reached \code{\$sqrt(NaN)}, a fatal
runtime math error that killed the simulation before any checker
ran. The fix guards non-finite $\hat v$ (NaN or negative $\to$ NaN,
$+\infty$ passes through), after which the reference predicts NaN
$m'/v'/w'$ for the poisoned lane and the NaN-aware scoreboard matches the
DUT's quiet-NaN outputs. Found by \code{dtx\_opt\_adamw\_test}. The lesson:
the reference is code too, and needs
adversarial inputs as much as the RTL does.

\emph{B2 (RTL): DMA write-error recovery re-latches the prior op's
error, a sticky-level vs.\ registered-clear race.} The AXI write engine's
\code{wr\_err} is a sticky \emph{level}, cleared only by \code{clr\_err};
the read engine's error is a self-clearing one-shot \emph{pulse}. The DMA
front end (\code{rtl/dtx\_dma.sv}) pulses \code{wr\_clr\_err} at op start,
but that pulse is registered and reaches the write engine one cycle into
the run state, while the front end's sticky capture
(\code{else if (op\_st \&\& wr\_err\_lvl)}) samples the error level on
exactly that first cycle, before the clear lands. A clean store issued
after a store that took \code{ERR\_AXI\_WR} therefore spuriously faulted
with the \emph{previous} op's error; read-error recovery, built on the
pulse idiom, was immune. The fix gates the capture with
\code{\&\& !wr\_clr\_err}: the suppressed cycle is exactly the one op-start
cycle on which the stale level is still visible, and no genuine B-channel
response can return that early, so no real error can be missed. Found by
\code{dtx\_dma\_test}'s fault-then-clean-rerun pattern, which
institutionalizes the audit rule that every fault test must
verify the reported outcome \emph{and} subsequent recovery. Mixing
level-sticky and pulse error idioms across two engines behind one
registered clear is the root cause worth generalizing.

\emph{B3 (methodology/architecture): naturally chained on-chip programs are
not verifiable against an exact-input oracle, and $dW$ needs a transpose
the chip cannot produce.} The first drafts of the acceptance tests chained
computed BF16 tiles directly into subsequent GEMMs on chip
(LN $\to$ cast $\to QK^{\mathsf T}$, noising $\to$ forward). Every
individual op was correct, yet the sweep blew up by a measured maximum
ratio of $5{,}340\times$ the Eq.~\eqref{eq:tol} budget. Root cause: the
golden model feeds each op its \emph{exact} FP64 reference input, while the
DUT forwards the BF16/PWL-rounded value; a ${\sim}2^{-8}$ input
perturbation lands in an output whose budget is ${\sim}2^{-20}$. Scaling
$s$ up by $5{,}000\times$ would ``fix'' the failure by verifying nothing.
Separately, the $dW = d\mathrm{pre}^{\mathsf T}x$ dataflow needs a
transposed operand that no on-chip primitive can produce, the DMA is a
pure row-major mirror with fixed inner stride. The resolution is the
deployment-realistic one: \emph{host-orchestrated stages}. Each stage
DMA-loads its BF16 GEMM operands from host memory (loads are always
exact-reference), computes, and stores; the host performs the BF16 casts
and the $d\mathrm{pre}^{\mathsf T}$ transpose between stages; FP32$\to$FP32
hops ($2^{-20}$ per hop) stay on chip within one program. Under this
structure both acceptance tests verify tight \emph{and} the loss falls.
Two v2 items follow directly: an on-chip transpose primitive (a
strided-inner DMA mode or a transposed GEMM drain) to un-host the $dW$
path, and a verifiability story for on-chip BF16 chaining (a golden-model
BF16 feed-forward mode, or per-hop tolerance re-baselining). A
test-authoring gotcha from the same debug is recorded with the fix:
tile-descriptor and memory-descriptor TDRs share the 16-entry file, and a
colliding index silently zeroes an operand with no scoreboard error (the
golden sees the same zeroed input), the convention is now tiles
0--7, memory descriptors 8--15.

The yield pattern is the section's summary: the numerics passed the
tolerance oracle essentially on arrival, the pipelined-primitive design
rules of Sec.~\ref{sec:numerics} were enforced by construction, and the
one real RTL bug lived in control-path error recovery, precisely where
byte-exact and tolerance methodologies look identical. What tolerance
verification \emph{cannot} see (a reference that shares the DUT's rounded
inputs, or mathematics that is consistently wrong on both sides) is exactly
what the semantic gates are for.

\providecommand{\synthNode}{sky130\,(130\,nm)}                 
\providecommand{\synthFmaxTT}{33.4\,MHz}                       
\providecommand{\synthFmaxSS}{7.4\,MHz}                        
\providecommand{\synthArea}{35.4\,mm\textsuperscript{2}}       
\providecommand{\synthCells}{2{,}576{,}969}                    
\providecommand{\synthPower}{3.45\,W}                          
\providecommand{\synthGflopsTT}{2.5\,GFLOP/s}                  
\providecommand{\synthGflopsSS}{0.56\,GFLOP/s}                 

\section{Results}\label{sec:results}

We report four tiers of numbers and keep them separate: (i) quantities
\emph{measured} in the 2026-07-07 RTL regression (simulation counters,
scoreboard statistics, coverage); (ii) per-cycle architectural accounting
read directly from parameters and verified latency contracts; (iii)
physical-implementation quantities measured by the 2026-08-09 sky130
synthesis campaign (Sec.~\ref{sec:results:synth}), each
labeled with the level at which it was measured (post-route with extracted
parasitics, or synthesized-netlist static timing); and (iv) clearly labeled
\emph{analytical} extrapolations.

\subsection{Regression outcome}\label{sec:results:regr}

Table~\ref{tab:regr} summarizes the full-suite regression of the 21-module,
8{,}170-line RTL. The headline is
not a single test but the conjunction: the same RTL that passes every
directed and control test also trains, the loss gate of
Sec.~\ref{sec:verif:tests} passes with a $2.2\times$ loss reduction while
every intermediate tensor of all 100 stage programs stays inside the derived
tolerance of Eq.~\eqref{eq:tol}, with zero failures among 107{,}108 checked
elements.

\begin{table}[t]
  \centering
  \caption{Full-suite regression, 2026-07-07 (all quantities measured).}
  \label{tab:regr}
  \footnotesize
  \begin{tabularx}{\columnwidth}{@{}X r@{}}
    \toprule
    Quantity & Value \\
    \midrule
    Tests passing & 17/17 \\
    \code{UVM\_ERROR} / \code{FATAL} / \code{WARNING} & 0 / 0 / 0 \\
    Scoreboard failures (all sweeps) & 0 \\
    \code{train\_mlp} programs replayed \& swept & 100 \\
    \code{train\_mlp} elements checked (exact + tol.) &
      107{,}108 (28{,}816 + 78{,}292) \\
    Training loss, iter.\ 0 $\to$ 19 & 56.44 $\to$ 26.02 \\
    Loss gate $\overline{\mathcal{L}}_{0..2}$ vs.\
      $\overline{\mathcal{L}}_{17..19}$ & 46.31 vs.\ 26.10 \\
    RNG mean / variance (98{,}304 samples) & 0.00812 / 1.00117 \\
    Overlap counters: GEMM + DMA vs.\ wall &
      $9{,}880 + 2{,}114 > 10{,}543$ \\
    PWL max err: \code{exp2} / GELU / GELU$'$ &
      $1.12{\times}10^{-5}$ (rel) / $2.21{\times}10^{-4}$ /
      $8.0{\times}10^{-4}$ (abs) \\
    Coverage: total / line / functional & 78.83 / 92.52 / 83.77\,\% \\
    \bottomrule
  \end{tabularx}
\end{table}

\subsection{Throughput accounting}\label{sec:results:flops}

Table~\ref{tab:flops} gives the peak per-cycle FLOP accounting, read from
the shipped parameters and the per-module latency/II contracts of
Sec.~\ref{sec:uarch}: the systolic array contributes $2\cdot\mathit{SYS\_N}^2$
(one exact BF16 multiply and one FP32 accumulate per PE per cycle at II=1),
the VPU one FP32 element per lane per cycle, and the fused-AdamW engine
$\sim$10 per lane per cycle. The optimizer figure is deliberately the
\emph{conservative} bookkeeping: a direct count of the AdamW update law is
16 semantic FLOPs per element, and the hardware physically executes
$\sim$37 FP-unit operations per element (including 12 rsqrt-Newton and 9
reciprocal-Newton ops); we book $\sim$10 and note that epilogue bias and
activation FLOPs during the GEMM drain are not counted at all. At the
simulation configuration this totals 216 FLOP/cycle, roughly $108\times$
the ${\sim}2$ FLOP/cycle of a loop-carried multiply--accumulate datapath
\emph{per clock}, before any frequency
advantage from the absence of loop-carried accumulation. The 128-lane
scale-out row is an architecture claim only: the VLIW/TDR control fabric
and tile-space bandwidth accounting scale with $\mathit{SYS\_N}$ and
$\mathit{LANES}$, but no RTL of that size has been built or timed.

\begin{table}[t]
  \centering
  \caption{Peak FLOP/cycle accounting. Sim and PD rows follow from shipped
  parameters and verified II=1 contracts; the scale-out row is analytical.}
  \label{tab:flops}
  \footnotesize
  \begin{tabularx}{\columnwidth}{@{}X r r r r r r@{}}
    \toprule
    Config & $\mathit{SYS\_N}$ & Lanes & GEMM & VPU & OPT & Total \\
    \midrule
    Sim / default            & 8   & 8   & 128    & 8   & 80    & \textbf{216} \\
    PD (\code{-DDTX\_SYS\_N=4 -DDTX\_LANES=4})& 4  & 4   & 32     & 4   & 40    & \textbf{76} \\
    Scale-out \emph{(analytical)} & 128 & 128 & 32{,}768 & 128 & 1{,}280 &
      \textbf{34{,}176} \\
    \bottomrule
  \end{tabularx}
\end{table}

Absolute throughput requires $f_\mathrm{max}$ and is deferred to
Sec.~\ref{sec:results:synth}; what simulation \emph{can} establish is that
the architecture sustains its peak, which is the next result.

\subsection{Compute--DMA overlap: sustaining the peak}
\label{sec:results:overlap}

\begin{figure*}[t]
  \centering
  \includegraphics[width=0.92\textwidth]{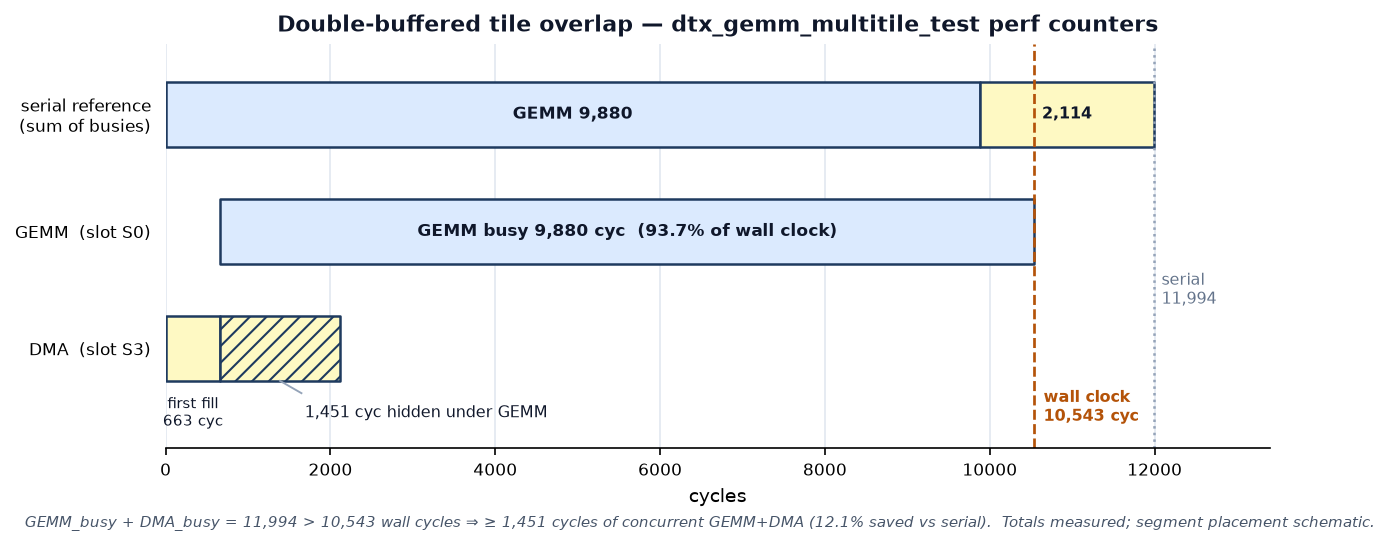}
  \caption{Double-buffered tile streaming measured by the performance
  counters of \code{dtx\_gemm\_multitile\_test}. Busy totals are measured;
  segment placement in the lower two rows is schematic.
  $\mathrm{GEMM\_busy}+\mathrm{DMA\_busy}=11{,}994 > 10{,}543$ wall cycles
  proves the overlap.}
  \label{fig:overlap}
\end{figure*}

Peak FLOP/cycle is worthless if the array starves between tiles.
\code{dtx\_gemm\_multitile\_test} runs the intended steady-state pattern: a
zero-overhead loop whose body co-issues a GEMM on one A-tile bank with a DMA
fill of the other (eight GEMM tiles, eight tile loads, RAW/WAR hazards
expressed as \code{wait\_mask} bits), measured with the on-chip performance
counters after \code{CNT\_CLR}. Over the 10{,}543-cycle program the GEMM
slot was busy 9{,}880 cycles, 93.7\% occupancy, and the DMA slot 2{,}114
cycles. The busy sum, 11{,}994, exceeds the wall clock by 1{,}451 cycles,
which is a counter-level \emph{proof} of concurrency: at least 1{,}451 DMA
cycles executed under GEMM compute, and the only exposed DMA is the
663-cycle first fill (Fig.~\ref{fig:overlap}). The ping-pong tile banks and
the VLIW co-issue model do what they were designed to do; a serial
schedule of the same work would be 13.8\% slower even on this short
program, and proportionally worse as loop counts grow.

\subsection{Training convergence: the semantic gate, measured}
\label{sec:results:train}

\begin{figure}[t]
  \centering
  \includegraphics[width=\columnwidth]{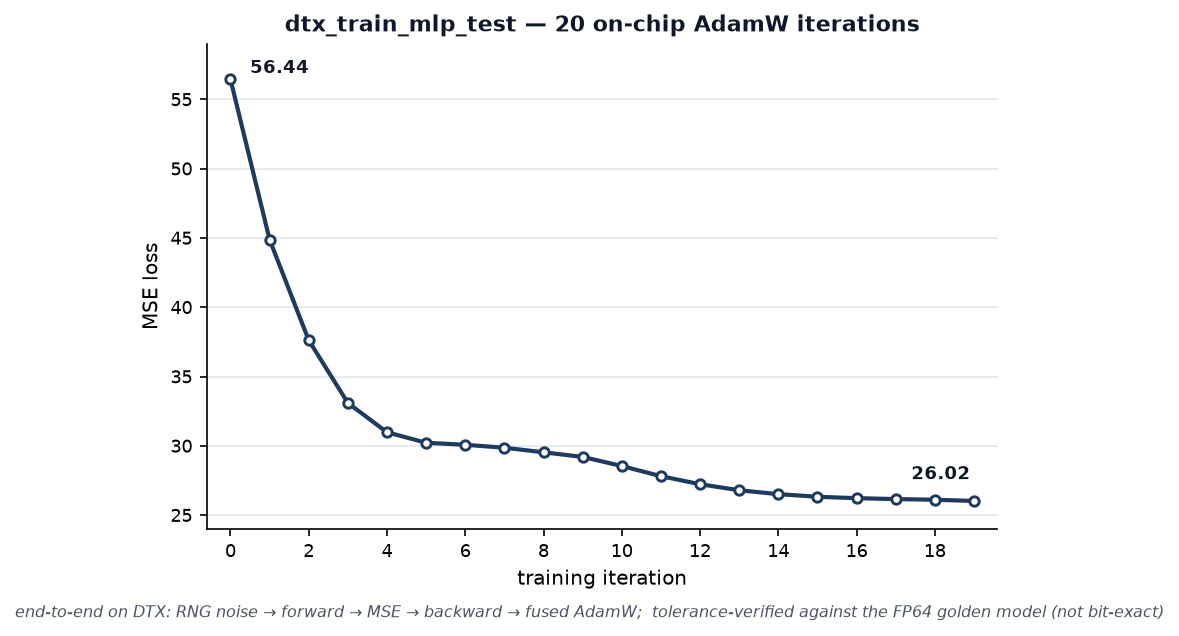}
  \caption{\code{dtx\_train\_mlp\_test}: MSE loss over 20 on-device
  training iterations (noise $\to$ forward $\to$ MSE $\to$ backward $\to$
  fused AdamW). The loss-decrease assertion is part of the regression;
  every intermediate tensor is simultaneously tolerance-checked against the
  FP64 golden model.}
  \label{fig:trainloss}
\end{figure}

Fig.~\ref{fig:trainloss} shows the twenty-iteration diffusion-MLP training
run of \code{dtx\_train\_mlp\_test} executing entirely on the device: per
step, a fresh (counter-reset, hence deterministic) Gaussian draw, closed-form
noising, forward GEMM with fused bias/ReLU, MSE gradient, activation
backward, $dW$ and $dX$ GEMMs, bias gradient, and two fused-AdamW updates,
with the weight held as an FP32 master and host-cast to BF16 for each
forward pass (the standard mixed-precision master-weight pattern). The MSE
loss falls monotonically from 56.44 to 26.02. Two properties make this a
result rather than a demo. First, it is an \emph{acceptance gate inside the
regression}: the test fails if the first-three-iteration mean loss does not
exceed the last-three mean, so any future change that silently breaks the
mathematics, while still matching a broken reference, is caught by the
workload itself (the shared-defect hazard of Sec.~\ref{sec:arch:csr},
guarded by a permanent regression artifact). Second, convergence here is not a substitute for numeric
checking but concurrent with it: every stored tensor of every stage program
in all twenty iterations is swept against the FP64 oracle. The companion
gate, \code{dtx\_train\_attn\_test}, closes the transformer side: a full
self-attention block forward and backward verifies clean through LayerNorm,
all three GEMM dataflows, softmax forward/backward, and the residual path.

\subsection{RNG quality}\label{sec:results:rng}

The Gaussian source is checked at two levels. Bit level: every Philox
uniform and every Gaussian draw in the directed RNG programs is compared
\code{EXACT} against the reference (the Philox datapath is integer; the
ICDF-LUT interpolation chain is mirrored operation-for-operation), and the
\code{RNG\_CTR} advance is scoreboard-predicted across ops \emph{and}
across program re-issues, so any draw is reproducible from
(seed, stream, counter). Statistical level: over 98{,}304 samples
($24$ re-issues of a $64\times 64$ Gaussian program, counter continuing so
every sample is fresh), the measured moments are mean $0.00812$ and
variance $1.00117$, within the test's bounds ($|\mu|<0.02$,
$|\sigma^2-1|<0.05$) and consistent with the LUT-law's own exact moments
(mean $+5.9\times10^{-3}$, variance $0.9973$, support $|z|\le 3.0973$ from
the 512-entry ICDF table), i.e.\ the sample statistics are explained by
the documented quantization of the method, not by a datapath defect.

\subsection{Iso-node comparison (analytical)}\label{sec:results:iso}

Everything in this subsection except the simulation counters already
reported and the per-cycle accounting of Table~\ref{tab:flops} is
\emph{analytical extrapolation}; none of it is silicon-measured. The claim
we defend is deliberately bounded: at iso-node and iso-area, a
fixed-function training pipeline of this shape should approach an order of
magnitude better throughput per watt than a general-purpose GPU stack, as
the product of two ratios.

\emph{Utilization ($\sim$2.0--2.6$\times$).} Published end-to-end
model-FLOPs utilization for large transformer training runs spans roughly
0.21--0.46 (GPT-3 21.3\%, Gopher 32.5\%, MT-NLG 30.2\%, PaLM
46.2\%~\cite{chowdhery2022palm}), with heavily optimized Megatron-style
stacks reporting $\sim$50\%~\cite{korthikanti2022megatron}; we take 0.35--0.46 as
the strong-baseline band. Weight-stationary systolic arrays sustain high
MAC utilization on dense training GEMMs by construction~\cite{jouppi2017tpu};
our in-simulation proxy is the 93.7\% GEMM-slot occupancy with overlapped
DMA of Sec.~\ref{sec:results:overlap}, a single-workload counter, not an
end-to-end MFU, so we assume 0.90. The ratio is $0.90/0.46\approx 2.0$
(conservative) to $0.90/0.35\approx 2.6$.

\emph{Energy per operation ($\sim$3--4$\times$).} DTX spends no energy on
warp scheduling, per-op instruction fetch/decode, or a large register file;
instruction and data supply dominate energy in general-purpose
pipelines~\cite{hameed2010inefficiency}, GPU power models attribute a double-digit
share of dynamic power to the register file and
pipeline~\cite{leng2013gpuwattch}, and the data-movement-versus-arithmetic
disparity is well established~\cite{horowitz2014energy}. The fused epilogue and
fused AdamW/LN/softmax additionally remove the elementwise-kernel DRAM
round-trips and launch overheads a GPU pays between GEMMs. We adopt
3--4$\times$ as the working band.

The product, $2.0\mbox{--}2.6 \times 3\mbox{--}4 \approx 6\mbox{--}10\times$,
approaches an order of magnitude, with $\sim$10$\times$ the favorable end.
The sky130 prototype exists to measure the one term simulation cannot:
FLOP/cycle/mm\textsuperscript{2} and energy at a real operating point; the
scaling argument beyond that node remains labeled as above.

\subsection{Synthesis and physical implementation (sky130, measured)}
\label{sec:results:synth}

DTX carries the verified RTL through a fully open flow:
sv2v $\to$ Yosys $\to$ OpenROAD via
SiliconCompiler on \synthNode~\cite{openroad2019,skywater2020}, at the PD
configuration \code{-DDTX\_SYS\_N=4 -DDTX\_LANES=4} (76 FLOP/cycle,
Table~\ref{tab:flops}), with both timing corners reported
(typical \code{tt\_025C\_1v80}; slow signoff
\code{ss\_n40C\_1v40}). The campaign ran on 2026-08-09; this section
reports what was measured and at which level. Throughout,
$f_\mathrm{max}=1/(T_0-\mathrm{WNS})$ under the $T_0=10$\,ns constraint
clock, with the SDC's 0.25\,ns clock uncertainty retained inside the
required time, so the quoted frequencies carry that margin.

\emph{The PD configuration is verified first.} Before synthesis, the full
17-test UVM regression was re-run at
$\mathit{SYS\_N}{=}4$, $\mathit{LANES}{=}4$: 17/17 green, including both
semantic gates. The run caught one defect, in the verification stimulus,
not the RTL: the error-injection sequence hardcoded its AdamW
base-misalignment fault as \code{0x10}, a $\mathit{LANES}{=}8$ constant.
The RTL's alignment rule is parameter-derived (legal bases satisfy
$\mathrm{base} \bmod (4\!\cdot\!\mathit{LANES}) = 0$), so at
$\mathit{LANES}{=}4$ the ``fault'' program was legal, issued, consumed four
never-loaded tiles, and the whole-space sweep flagged 194 X-valued
elements, the X-poisoning of unwritten tile SRAM doing exactly its job.
The stimulus now computes its violating base from the same parameter the
RTL derives the rule from (half the $4\!\cdot\!\mathit{LANES}$ alignment
granule: FP32-word-aligned, never granule-aligned for any
$\mathit{LANES}\ge 2$), and both configurations re-ran green. The overlap proof of
Sec.~\ref{sec:results:overlap} also holds at the PD configuration:
$31{,}256$ GEMM-busy $+$ $2{,}121$ DMA-busy $>$ $31{,}925$ wall cycles on
the same eight-tile program (97.9\% GEMM-slot occupancy, higher than the
simulation configuration's 93.7\% because $8{\times}8$-dim tiles take four
passes on the $4{\times}4$ array, adding compute per DMA byte).

\emph{Synthesis: the design law held; the memories are the cost.} The full
chip maps cleanly through hierarchical Yosys synthesis in 56 minutes:
\synthCells{} standard cells, \synthArea{} of cell area
(Table~\ref{tab:synth}). Two findings frame that number. First, ABC maps
every FP cone in
minutes: the pipelined-FP design law of Sec.~\ref{sec:numerics} did at
full-chip scale exactly what it was written to do (the one flow adjustment
was disabling \emph{flattened} synthesis, whose SAT-based resource-sharing
pass, not ABC, exhausted memory on the 2.6M-cell netlist). Second, the
area is dominated by the open flow's lack of matching SRAM macros: the
three FF-mapped regions of the 64\,KB tile space are
24.6\,mm\textsuperscript{2} (region~C, 32\,KB, 12.3; A/B, 16\,KB each, 6.2
apiece; 694k cells for C alone) and the IRAM-bearing CSR block another
2.7, roughly 77\% of all cell area is architectural memory held in
flip-flops. The compute engines are small by comparison (synthesis cell
area within the full chip): fused-AdamW 4.7\,mm\textsuperscript{2}, VPU
2.0, RNG 0.7 (Philox 0.4 within it), systolic array 0.31, epilogue 0.23.

\emph{Full-chip timing and power (synthesized netlist).} Static timing on
the mapped netlist (zero-wire-load, both corners) gives
$f_\mathrm{max}=$ \synthFmaxTT{} typical and \synthFmaxSS{} slow; the
critical path is \emph{not} an FP cone but the write network into a tile
region (endpoint: a region-B memory-bit flop), the FF-mapped memory again,
not the datapath. Power at the same netlist, typical corner, 100\,MHz, is \synthPower{},
using a \emph{uniform global} switching activity of 0.045
toggles/bit/cycle, the chip-average measured from the PD-configuration
double-buffering workload's RTL VCD over its 31{,}925-cycle busy window
(whole-test average 0.049) and applied identically to every net; 94\% of
the total is sequential internal power, the clock power of $\sim$600k
memory-bit flops with no clock gating. Leakage is 12\,\textmu W, so
dynamic-only frequency scaling is exact: $\approx$1.2\,W at the 33.4\,MHz
netlist-level operating point (an upper bound, the zero-wire-load netlist
accounts flop-internal clock power but no clock tree or wire capacitance).
These are pre-layout numbers and labeled as such; the post-route engine
data below anchors the wire-load reality.

\begin{figure}[t]
  \centering
  \includegraphics[width=0.86\columnwidth]{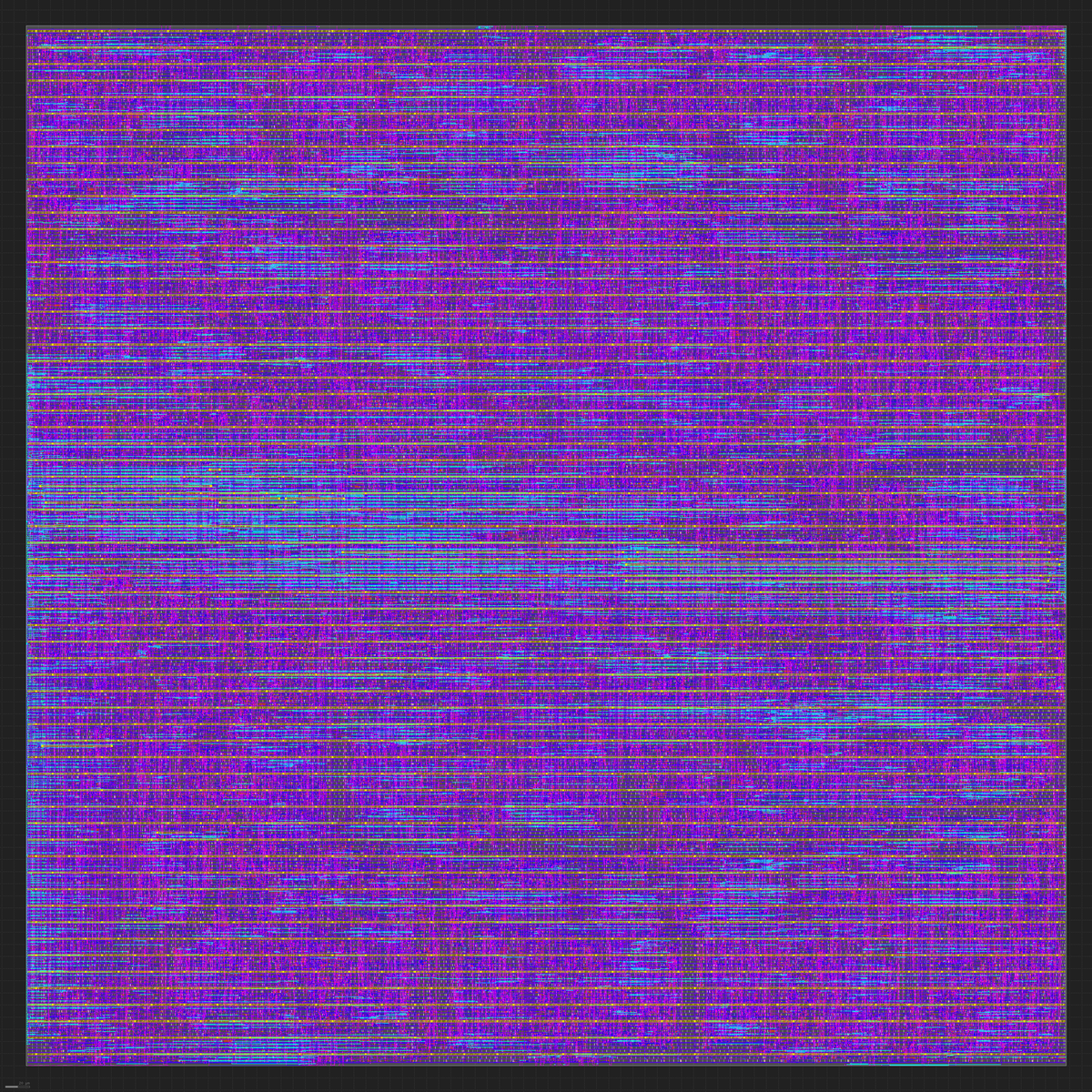}
  \caption{Routed \code{dtx\_gemm\_array} (sky130hd, 0.73\,mm\textsuperscript{2}
  core at 52\% utilization, 0 DRC violations): the $4{\times}4$
  weight-stationary array hardened standalone by the OpenROAD flow.}
  \label{fig:die}
\end{figure}

\emph{Engine-level place-and-route (post-route, extracted parasitics).}
Every compute-engine class was hardened standalone through the complete
flow, and all five close to router-DRC-clean GDS with hold met at both
corners (Table~\ref{tab:engines}; Fig.~\ref{fig:die} shows the routed
array). ``DRC'' here means detailed-route
violations (TritonRoute); antenna checking is disabled in this
characterization flow and no independent signoff DRC deck was run. The
headline is the systolic array: detailed route converges with \emph{zero}
router violations, hold met at both corners, and post-route
$f_\mathrm{max}$ of 83.3\,MHz typical / 19.8\,MHz slow, with the critical
path ending at a PE accumulator's \code{fp32\_add} stage register, i.e.\
the machine sits exactly on the pipelined-FP limit the architecture
intended. As a calibration point on the same node and flow, a
training datapath built around a loop-carried MAC that we hardened
through the identical toolchain
(full-chip post-route, sky130hd) was capped at 45\,MHz typical by its
serial \code{fp32\_add} even after three rounds of pipelining
optimization around the chain; the DTX array's post-route 83.3\,MHz is
$1.9\times$ that,
measured at the same level, and the II=1 tree/systolic structure needs no
such rescue. The VPU, Philox, and epilogue close the same way, router-DRC
clean, hold met at both corners, 76--85\,MHz typical, and every one of
their critical paths ends at a pipeline-stage register of one FP or
Philox-round unit, i.e.\ the whole engine fleet sits on the intended
one-op-per-stage limit. The fused-AdamW engine congests at 40\% placement density
(its 4-lane $m/v/g/w$ streaming fan-in) and closes at 25\%: router-DRC-clean GDS, hold
met at both corners, 76.9\,MHz typical / 18.7\,MHz slow, with the critical
path one \code{fp32\_add} stage in the weight-finalize adder, roughly 37
physical FP operations per element, every one of them exactly one pipeline
stage deep.

\emph{What full-chip P\&R needs, stated plainly.} Full-chip
place-and-route was not attempted to completion: 2.6M
instances, 1.6M of them in the FF-mapped memory blocks, including the
$\sim$600k memory-bit flops and their read/write-mux networks, is
roughly five times the largest design this open flow has closed on the
30\,GB build machine (a 536k-instance inference chip). The enabler
is the SRAM macro path: each 2R1W tile region maps onto
write-through-duplicated 1RW1R macro pairs, replacing
$\sim$27\,mm\textsuperscript{2} of flops and their write/read networks
(and, per the full-chip critical path above, the frequency limiter) with
compiled arrays. That iteration is scoped in Sec.~\ref{sec:conclusion};
what this campaign establishes is that the full chip synthesizes with
every FP operator, Newton units included, mapped as real logic, and that
every hardened engine's critical path lands on a pipeline-stage register
of a single FP or Philox-round unit, the design sitting exactly on its
intended one-op-per-stage limit.

\begin{table}[t]
  \centering
  \caption{Full-chip physical summary (\synthNode, PD configuration
  $\mathit{SYS\_N}{=}4$, $\mathit{LANES}{=}4$). Netlist-level quantities
  (hierarchical synthesis; zero-wire-load STA); measured activity 0.045
  toggles/bit/cycle.}
  \label{tab:synth}
  \footnotesize
  \begin{tabularx}{\columnwidth}{@{}X r@{}}
    \toprule
    Quantity & Value \\
    \midrule
    Standard cells                    & \synthCells \\
    Cell area (77\% FF-mapped memory) & \synthArea \\
    $f_\mathrm{max}$, typical corner (netlist)  & \synthFmaxTT \\
    $f_\mathrm{max}$, slow corner (netlist)     & \synthFmaxSS \\
    Critical path & tile-region write network \\
    Power @ 100\,MHz, tt, act.\ 0.045 & \synthPower \\
    Power @ 33.4\,MHz (netlist)       & $\approx$1.2\,W \\
    Peak throughput, typical (76 FLOP/cycle $\times f_\mathrm{max}$)
                                      & \synthGflopsTT \\
    Peak throughput, slow             & \synthGflopsSS \\
    \bottomrule
  \end{tabularx}
\end{table}

\begin{table}[t]
  \centering
  \caption{Engine-level place-and-route (post-route STA with OpenRCX
  parasitics; $f_\mathrm{max}$ from worst setup slack). Core area $=$
  post-route cell area $/$ placement utilization (52--54\%; AdamW closed at
  25\% target density after congesting at 40\%, hence 34\% and the roomier
  core); the smaller figures in the text are synthesis cell areas within
  the full chip. DRC $=$ detailed-router violations. Power at typical
  corner, uniform global activity 0.045, scaled to the tt
  $f_\mathrm{max}$.}
  \label{tab:engines}
  \footnotesize
  \begin{tabularx}{\columnwidth}{@{}X r r r r r@{}}
    \toprule
    Engine & Core (mm\textsuperscript{2}) & DRC &
    $f_\mathrm{max}$ tt/ss (MHz) & Hold & P (mW) \\
    \midrule
    GEMM array & 0.73 & 0 & 83.3 / 19.8 & met & 28 \\
    VPU        & 4.75 & 0 & 76.2 / 18.2 & met & 145 \\
    Philox     & 0.99 & 0 & 80.0 / 20.4 & met & 16 \\
    Epilogue   & 0.55 & 0 & 85.2 / 20.2 & met & 22 \\
    AdamW      & 14.6 & 0 & 76.9 / 18.7 & met & 308 \\
    \bottomrule
  \end{tabularx}
\end{table}

\section{Related Work}\label{sec:related}

\textbf{Systolic and dataflow DNN accelerators.} The TPU~\cite{jouppi2017tpu}
established the weight-stationary systolic array as the workhorse for dense
neural GEMMs, and Eyeriss~\cite{chen2016eyeriss} mapped the wider dataflow
design space; roofline analysis~\cite{williams2009roofline} frames the
bandwidth-versus-compute trade that DTX's ping-pong tile buffers and 2-D
burst DMA address. DTX's array follows the TPU lineage at tile scale
($8{\times}8$, dims to 64 via multi-pass with FP32 partial-sum round-trips
through the output SRAM) but is training-specific: three dataflows including
accumulate-onto-memory $dW$, an FP32 psum path with a fused
bias/activation/cast drain epilogue, and BF16-in/FP32-accumulate
numerics~\cite{kalamkar2019bf16,micikevicius2018mixed} throughout.

\textbf{VLIW and EPIC.} Statically scheduled wide issue originates with
VLIW~\cite{fisher1983vliw} and matured into the EPIC
philosophy~\cite{schlansker2000epic}, explicit dependence information
carried in the instruction so the hardware need not rediscover it, %
commercially embodied in DSPs such as TI's C6000
family~\cite{ti2000c6000}. DTX is EPIC at engine granularity rather than
functional-unit granularity: its four slots name whole engines (GEMM,
vector, optimizer/RNG, DMA), its \code{wait\_mask} bits are inter-word drain
dependences honored by a scoreboard, and variable DMA latency is absorbed by
drain semantics rather than by the compiler's cycle model. The
zero-overhead loop is likewise standard DSP practice. What DTX takes from
this literature is the division of labor, dumb fast hardware, smart
compiler, justified here by a workload (the training step) that is a
compile-time-known DAG of fixed-latency tile ops.

\textbf{GPU training baselines.} Our iso-node comparison
(Sec.~\ref{sec:results}) is anchored to published end-to-end
model-FLOPs-utilization for large-transformer training, 21--46\% across
GPT-3, Gopher, MT-NLG, and PaLM~\cite{chowdhery2022palm}, with heavily
optimized stacks reaching ${\sim}50\%$~\cite{korthikanti2022megatron}, and
to the energy accounting of general-purpose pipelines: instruction and data
supply dominating datapath energy~\cite{hameed2010inefficiency}, register
file and scheduling as a double-digit share of GPU dynamic
power~\cite{leng2013gpuwattch}, and the arithmetic-versus-movement energy
disparity~\cite{horowitz2014energy}. We use these sources to bound the two
factors of the claim (utilization $2.0$--$2.6\times$, energy per op
$3$--$4\times$) and present the product as analytical, not measured.

\textbf{Reproducible and tolerance-based numerics.} Counter-based RNGs were
designed for exactly the replay property DTX
exploits~\cite{salmon2011philox}; DTX checks the Philox stream bit-exactly at
the integer level and the Gaussian transform statistically. On the summation
side, the error analyses of sequential versus pairwise
reduction~\cite{higham1993summation,goldberg1991floating} supply the
$\log_2 K$ scaling of our tolerance bound; Demmel and
Nguyen~\cite{demmel2013reproducible} show the cost of restoring
reproducibility to parallel sums in software, and NVIDIA's own
guidance~\cite{whitehead2011precision} documents that GPU reductions are not
bit-reproducible across configurations and must be compared under
tolerance, the same conclusion our hardware contract reaches, but stated
here normatively, with the bound derived from the datapath's actual
reduction depths and validated for tightness (Sec.~\ref{sec:verif}).

\section{Conclusion}\label{sec:conclusion}

DTX demonstrates that a training accelerator can be built with no
loop-carried arithmetic anywhere, systolic
GEMM, tree reductions, fully pipelined FP with initiation interval~1
throughout, and reach 216~FLOP per cycle of peak in the simulated
configuration, with compiler-scheduled 4-slot VLIW control whose overlap is
proven, not assumed: 9{,}880 GEMM-busy plus 2{,}114 DMA-busy cycles against a
10{,}543-cycle program. The verification contract matches the
architecture, deliberately: an FP64 golden model under a tolerance derived
from the datapath's reduction depths, exact equality where the math is
exact, and a semantic oracle in the regression, a diffusion-MLP training
run whose loss falls 56.4 to 26.0 over 20 in-testbench optimizer steps, plus
an attention block verified forward and backward. Seventeen of seventeen
tests pass with 107{,}108 elements checked and zero scoreboard failures; the
bug catalog holds three defects, exactly one of them in the RTL (a
sticky-level versus registered-clear race in DMA write-error recovery, the
kind of cross-module protocol bug that survives unit thinking and falls to a
fault-then-recover test). The sky130 campaign then attached the first
physical numbers: the PD configuration passes the same 17-test regression,
the full chip synthesizes with every FP operator mapped as real logic,
and all five engine classes, including the fused AdamW, close to
router-DRC-clean GDS at 76.9--85.2\,MHz post-route typical, the systolic array's 83.3\,MHz
$1.9\times$ the post-route ceiling of an optimized loop-carried
multiply--accumulate baseline on the same node and flow
(Sec.~\ref{sec:results:synth}).

The methodological finding is symmetric to the architectural one: a
tolerance bound certifies nothing outside its premises. The bound was tight
enough that the one program which broke its premise, chaining a computed
BF16 tile directly into a GEMM whose reference consumed the exact FP32
value, failed by $5{,}340\times$, and the resolution was architectural
honesty rather than tolerance inflation: end-to-end training runs as
host-orchestrated stages, each stage DMA-loading exact-reference BF16
operands, which is precisely how a host driver launches kernels in
deployment. The verification contract is part of the architecture, and the
regression ends where it must: with a real workload as the
final oracle.

v2 work is scoped by what v1 measured. First, an on-chip transpose
primitive, a strided-inner DMA mode or a transposed GEMM drain, so the
$dW = dY^{\mathsf T}X$ operand no longer requires the host, the one gap that
forces staging today. Second, a golden-model strategy for on-chip BF16
chaining, a BF16 feed-forward reference mode or per-hop tolerance
re-baselining, so naturally chained programs verify without host
round-trips. Third, pass-overlap in the GEMM engine: draining pass $p$ while
loading the weights of pass $p{+}1$, closing most of the gap between the
measured 93.7\% slot occupancy and the array's streaming bound. Fourth,
coverage closure: the first-pass merged figure is 78.83\% total (92.52\%
line, 83.77\% functional), and the strict-exclusion flow with authored
waiver files must precede any closure claim. Fifth, from the measured synthesis
campaign (Sec.~\ref{sec:results:synth}): the SRAM-macro tile space
(write-through-duplicated 1RW1R pairs per 2R1W region), the change that
turns a full chip which already \emph{synthesizes} at 2.58M cells, 77\% of
whose cell area is FF-mapped memory, into a routable die, removes the
memory-write-network critical path that caps the netlist at 33.4\,MHz, and
would let chip-level frequency approach the engine fleet's measured
76--85\,MHz post-route band, subject to full-chip routing and clock
distribution, which no configuration of this design has yet measured.

\bibliographystyle{IEEEtran}
\bibliography{refs}

\end{document}